\documentclass[12pt]{article}
\usepackage{epsfig, epsf, graphicx, subfigure, color}
\usepackage{pstricks, pst-node, psfrag}
\usepackage{amssymb,amsmath,bm}
\usepackage{amsmath}
\usepackage{verbatim,enumerate}
\usepackage{rotating, lscape}
\usepackage{dsfont}
\usepackage{setspace}
\usepackage{bbm}
\usepackage{amsmath}
\usepackage{natbib}
\usepackage{amsthm}
\usepackage{enumerate}
\usepackage{url}
\usepackage{appendix}
\usepackage{amssymb}
\usepackage{tabularx,booktabs}
\usepackage{amsfonts}
\DeclareMathAlphabet{\mathcal}{OMS}{cmsy}{m}{n}
\usepackage[T1]{fontenc}
\usepackage{float}
\usepackage{dsfont}
\usepackage{tabularx}
\usepackage{soul}
\usepackage{multirow}
\usepackage[cal=cm]{mathalfa}
\graphicspath{{plots/}}
\usepackage[pdftex,colorlinks=true]{hyperref}
\usepackage{mathalfa}
\definecolor{darkblue}{rgb}{0,0,.6}
\hypersetup{citecolor=darkblue,linkcolor=darkblue,urlcolor=darkblue}
\usepackage{tabularx}
\graphicspath{{plots/}}

\theoremstyle{definition}
	
\newtheorem{thm}{Theorem}
\newtheorem{dfn}[thm]{Definition}

\newcommand{\bi}{\begin{itemize}}
	\newcommand{\ei}{\end{itemize}}

\begin{document}
	\thispagestyle{empty} \baselineskip=28pt \vskip 5mm
	\begin{center} {\Large{\bf Pointwise Data Depth for Univariate and Multivariate Functional Outlier Detection}}
	\end{center}
	
	\baselineskip=12pt \vskip 5mm
	\begin{center}\large
		Cristian F. Jim\'enez-Var\'on\footnote[1]{
			\baselineskip=10pt CEMSE Division, Statistics Program, King Abdullah University of Science and Technology, Thuwal 23955-6900, Saudi Arabia. E-mail: cristian.jimenezvaron@kaust.edu.sa; fouzi.harrou@kaust.edu.sa, ying.sun@kaust.edu.sa}, Fouzi Harrou$^1$, Ying Sun$^1$\end{center}	
	
	\baselineskip=16pt \vskip 1mm \centerline{\today} \vskip 8mm
	
	\begin{center}
		{\large{\bf Abstract}}
	\end{center}
	\baselineskip=17pt
 Data depth is an efficient tool for robustly summarizing the distribution of functional data and detecting potential magnitude and shape outliers. Commonly used functional data depth notions, such as the modified band depth and extremal depth, are estimated from pointwise depth for each observed functional observation. However, these techniques require calculating one single depth value for each functional observation,  which may not be sufficient to characterize the distribution of the functional data and detect potential outliers. This paper presents an innovative approach to make the best use of pointwise depth. We propose using the pointwise depth distribution for magnitude outlier visualization and the correlation between pairwise depth for shape outlier detection. Furthermore, a bootstrap-based testing procedure has been introduced for the correlation to test whether there is any shape outlier. The proposed univariate methods are then extended to bivariate functional data. The performance of the proposed methods is examined and compared to conventional outlier detection techniques by intensive simulation studies. In addition, the developed methods are applied to simulated solar energy datasets from a photovoltaic system. Results revealed that the proposed method offers superior detection performance over conventional techniques. These findings will benefit engineers and practitioners in monitoring photovoltaic systems by detecting unnoticed anomalies and outliers.
	\begin{doublespace}
		
		\par\vfill\noindent
		{\bf Some key words}: Data depth; Functional data; Magnitude outliers; Pairwise depth; Pointwise depth; Shape outliers; Visualization.
	\par\medskip\noindent
		{\bf Short title}: Pointwise Data Depth for Functional Outlier Detection.
	\end{doublespace}
	
	\pagenumbering{arabic}
 
\section{Introduction}\label{sec:intro}

The number of countries implementing renewable energy policies increased again in 2021, maintaining a multi-year trend, with solar photovoltaic (PV) being the most popular renewable energy performer \citep{REN21}. Because of variable climatic factors such as cloud cover and daily temperature, power generation from PV systems may not be consistent. This fluctuation can seriously affect PV systems; for example, short-term variations (a few seconds) might cause local voltage flickering, and a longer-term time scale change can impact the storage system \citep{Kleissl2013}. As a result, improper detection of such fluctuations in daily performance might result in many disruptions and damages to PV production \citep{Mallor2017,Euan2022}.

Outlier observations of PV power and related covariables, such as solar irradiance, can be statistically examined using a variety of methodologies \citep{Marcos2011,Solorzano2013,Garoud2017}. Among such approaches, functional data analysis (FDA) has provided a variety of outlier detection strategies in environmental, meteorological, medical, and economic contexts over the last two decades \citep{Febrero2008,ullah2013}. FDA considers each observation as  a function defined over a continuous interval. \cite{Ramsay} book is one of the founding works on functional data analysis. It provides various parametric methods for modeling and inference. Based on the fact that functional data is intended to be continuous, various non-parametric smoothing methods for functional data are proposed in \cite{Ferraty}. Essentially, functional data observations are considered as realizations of a stochastic process.  In the FDA literature, two types of anomalous observations are usually considered: magnitude and shape outliers \citep{Hubert2015,TVD,MBD}. Data from PV systems with aged sensors, for example, could be considered shape outliers, whereas data from PV systems in dusty conditions could be considered magnitude outliers \citep{Garoud2017}.

One common way of describing the data distribution is through order statistics. In particular, ranks and quantiles have been widely used for statistical inference. However, in multivariate settings, the concept of order is not as natural as in the univariate case \citep{tukey,Barnett1976}. Data depth has been studied intensively in non-parametric statistics \citep{Zuo2000}. It has a close relation to multivariate ranks and outlier detection. The depth function indicates a center-outward ordering of multivariate functional data \citep{Liu1999}. Some common notions of depth for multivariate data include~\cite{mahalanobis,tukey,simpl,Zuo2003}. In terms of univariate functional data, numerous methodologies have introduced the concept of functional depth over the past two decades; for instance, the modified band depth (MBD) measures the proportion of time a curve is in a random band \citep{MBD}. The integrated data depth defined by \cite{Fraiman} proposes to use an integrated version of the pointwise depth over the interval $[0,1]$. Some other notions include~\cite {Cuevas,halfregion,Narisetty2016,TVD}.  For multivariate functional data, there exist other depth notions, such as functional Tukey depth proposed by \cite{Dutta}, functional spatial depth introduced by \cite{Chakraborty}, and the directional outlyingness and the MS plot proposed in \cite{Dai2018,msplot}.

Functional data applications often exhibit abnormal observations, i.e.,  magnitude and shape outliers.  \cite{Sun2011} proposed the functional boxplot, in which the potential outliers are detected by the 1.5 times $50\%$ central region rule, similar to the classical boxplot. The functional boxplot is effective in detecting magnitude outliers. This is because its outlier detection rule examines how far functional observations deviate from the established $50\%$ central region. Existing outlier detection methods typically remove the magnitude outliers flagged by the functional boxplot before finding shape outliers \citep{outliergram, TVD}. To detect shape outliers, \cite{outliergram} proposed another visualization tool, "the outliergram" (OG). Its outlier detection rule uses the plane formed by  MBD and modified epigraph index (MEI). Any observations that are beyond the parabola constructed from the approximated mathematical relationship of  MBD and MEI are flagged as shape outliers. \cite{Dai2018} proposed a graphical tool, the magnitude-shape MS plot. This tool allows the visualization of both the magnitude and shape outliers in functional data. The MS plot is built on the notion of functional directional outlyingness defined by \cite{msplot}, which measures the centrality of functional data by simultaneously considering the level and the direction of their deviation from the central region. \cite{TVD} proposed the concept of modified shape similarity (MSS) for shape outlier detection, derived from the decomposition of the proposed total variation depth (TVD). While the MS plot is developed for multivariate functional data, the MSS is only for univariate cases.

Each of the aforementioned outlier detection tools has its advantages and disadvantages. The main objective of this paper is to propose a set of functional data visualization and outlier detection tools based on pointwise data depth. We are motivated by the definition of MBD, TVD, and the extremal depth (ED) \citep{Narisetty2016}. The sample depth value for each functional observation is computed via pointwise depth for  MBD, TVD, and ED notions. In particular, MBD and TVD average the pointwise depth values. The modified simplicial band depth introduced by \cite{multi},  defines the depth notion based on the average of the pointwise simplicial depth. We propose to visualize the distribution of the pointwise depth for each functional data to identify potential purely or partially magnitude outliers. Then, we use the correlation between two consecutive pointwise depth values as a measure for shape outlyingness. Moreover, most existing shape outlier detection methods are developed for data with contaminations and tend to flag false shape outliers when no outliers are present \citep{outliergram,TVD,msplot}. Therefore, we propose a hypothesis testing procedure for the existence of shape outliers. Finally, we show the extension of our proposed method to the multivariate case. We demonstrated the ability of the proposed outlier detection approach compared to commonly used outlier detection methods using univariate and multivariate synthetic data. In addition, we applied our approach to measurements collected from a simulated PV system. These measurements include solar irradiance and the generated PV power. Monitoring PV systems for outlier detection can provide early indications of anomalies, which is required to maintain these systems under the expected performance. Results showed the ability of the proposed approach to uncover some outliers that are undetectable by the conventional methods.

The rest of the  paper is organized as follows. In Section~\ref{sec:method}, we present a methodology for using the pointwise data depth for functional data outlier detection. We illustrate how the pointwise depth and the pairwise pointwise depth are used to characterize the outlyingness in univariate and bivariate functional data. We also present a bootstrap-based procedure to test for the existence of shape outliers. The simulation study is performed in  Section~\ref{sec:Simulation}, and the applications for univariate and multivariate functional data are illustrated on a dataset from a photovoltaic system, where daily curves of PV power and solar irradiance from multiple years are considered in Section~\ref{sec:application}. We conclude and discuss the paper in Section~\ref{sec:Discussion}. 

\section{Methodology}\label{sec:method}

\subsection{Pointwise data depth for functional data}\label{2.1}

The pointwise data depth is a depth measure proposed to best characterize the distribution of functional data evaluated at discrete time points. We will illustrate how to use the pointwise data depth (PWD) for magnitude outlier visualization and shape outlier detection for functional data.

Let $X$ be a real-valued stochastic process index by $\mathcal{T}$ with distribution function $F_X \in \mathbb{R}$.   We propose the pointwise depth for a given function $f$ with respect to $F_X$. First, Let $R_f(t)=\mathds{1}\{X(t) \leq f(t)\}$, where $\mathds{1}$ is the indicator function. We can observe that   $p_f(t):=\mathds{E}[R_f(t)]=\mathds{P}\{X(t) \leq f(t)\}$ is associated with the relative position of $f(t)$ with respect to $X(t)$. For a given function $f(t)$ at each fixed $t$, we define the pointwise depth of $f(t)$ as

 \begin{equation}
     \text{PWD}_f(t;F_X)= 2\text{var}\{R_f(t)\}=2p_f(t)\{1-p_f(t)\}.
     \label{pop_PWD}
 \end{equation}

\noindent  For each fixed $t$, $\text{PWD}_f(t;F_X)$ is maximized at the center when $p_f(t)=1/2$, the univariate median, and the maximum $\text{PWD}_f(t;F_X)$ is $1/2$. When data are observed, the $\text{PWD}_f(t;F_X)$ can be estimated by approximating $\text{PWD}_f(t;F_X)$ by pointwise bands. In the sample version, consider $S:=\{X_1(t),\ldots,X_n(t)\}$ be a set of realizations of $X$, in which each $X_i(t)$ is assumed to be independent and measured at $t_j,j=1,\ldots,p$, in $\mathcal{T}$. Let $f(t)$ be a given function, a realization of $X$ that may or may not be a member of $S$. For each fixed $t \in \mathcal{T}$, define the sample pointwise depth of $f$ with respect to $S$ as

 \begin{equation*}
\widehat{\text{PWD}}_f(t_j;S)=\frac{n_a\times n_b +n-1}{{n \choose 2}},
\end{equation*}

\noindent where, $n_a$ and $n_b$ denotes the number of points above/below $X_i(t_j)$ at time point $t_j$ respectively. This estimator is a special case of the modified band depth (MBD) when the number of time points is 1.

\noindent Definitions \ref{MBD} and \ref{TVD}, introduce the population definition and sample versions of MBD by \cite{MBD}, and  TVD by \cite{TVD}. Later, we show the relationship of the sample versions of MBD and TVD with $ \widehat{\text{PWD}}$.

\begin{dfn}\label{MBD} \textbf{Band Depth (BD) and modified band depth (MBD)} 

\cite{MBD} introduces the band depth concept through a graph-based approach. The graph of a function $f(t)$ is the subset of the plane $G(f)=\{(t, f(t)):t \in \mathcal{T}\}$. The band in $\mathds{R}^2$ delimited the curves $x_{i1},\ldots,x_{ik}$ is $B(x_{i_1},\ldots,x_{i_k})=\{(t,f(t)):t \in \mathcal{T},\min_{r=1,\ldots,k}x_{i_r}(t)\leq f(t) \leq \max_{r=1,\ldots,k} x_{i_r}(t)\}$. Let $J$ be the number of curves determining a band, $2\leq J \leq n$. If $X_1(t),\ldots, X_n(t)$ are independent realizations of the stochastic process $X(t)$ generating the observations $x_1(t),\ldots,x_n(t)$, the population version of the band depth for a given curve $f(t)$ with respect to the probability measure $P$ is defined as follows:
\begin{equation*}
    \text{BD}_J(f;P)=\sum_{J=2}^J\text{BD}^{(J)}(f,P)=\sum_{j=2}^JP\{G(f) \subset B(X_1,\ldots,X_j)\}
\end{equation*}
The sample version of $\text{BD}_J(f; P)$ is computed by computing the fraction of the bands determined by $j$ different sample curves containing the whole graph of the curve $f(t)$, that is, $\text{BD}_n^{(j)}(f)={n\choose j}^{-1}\sum_{1\leq i_1<i_2<\cdots<i_j\leq n}\mathbbm{1}\{G(f) \subseteq B(x_{i_1},\ldots,x_{y_{i_j}})\}$, where $\mathbbm{1}\{\cdot\}$ denotes the indicator function. The MBD, also proposed by \cite{MBD} is a more flexible definition that measures the proportion of time that a curve $f(t)$ is in the band. The MBD is defined as 
\begin{equation*}
    \text{MBD}^{(j)}_n(f)={n\choose j}^{-1}\sum_{1\leq i_1<i_2<\cdots<i_j\leq n}\lambda_r\{A(f;x_{i_1},\ldots,x_{i_j})\},
\end{equation*}
where $A_j(f)\equiv A(f;x_{i_1},\ldots,x_{i_j}) \equiv \{ t \in \mathcal{T}: \min_{r=i_1,\ldots,i_j} x_r(t) \leq f(t) \leq \max_{r=i_1,\ldots,i_j} x_r(t)\}$ and $\lambda_r(f)=\lambda(A_j(f))/\lambda(\mathcal{T})$, if $\lambda$ is the Lebesgue measure on $\mathcal{T}$.

\end{dfn}
It is clear that the proposed pointwise depth, in its sample version, has direct connection to the sample version of the MBD.

\begin{equation*}
    \text{MBD}_{n}^{(2)}(f)=\frac{1}{p}\sum_{j=1}^p \widehat{\text{PWD}}_{f}(t_j;S)
\end{equation*}

\begin{dfn}\label{TVD} \textbf{Total variation depth (TVD)}

Let X be a real stochastic process on $\mathcal{T}$, with distribution $F_X$, where $\mathcal{T}$ is an interval in $\mathds{R}$. For a function $f(t)$, let $R(t)$ be the indicator function $R_f(t)=\mathds{1}\{X(t) \leq f(t)\}$ for $t\in \mathcal{T}$. The functional total variation depth of \cite{TVD} of the function $f(t)$ with respect to $F_X$ is defined as
\begin{equation*}
    \text{TVD}(X;F_X)=\int_{\mathcal{T}}D_X(t)\omega(t)dt,
\end{equation*}
where $\omega(t)$ is a weight function and $D_X(t)$ is the pointwise total variation depth defined by,
\begin{equation*}
    D_X(t)=\text{var}\{R_f(t)\}=\mathds{P}(X(t)\leq f(t))\mathds{P}(X(t)>f(t)).
\end{equation*}
   
\end{dfn}

Although TVD and MBD have slightly different definitions, they are equivalent when it comes to establishing the order of data.

\begin{figure}[!ht]
\centering
\includegraphics[width=70mm]{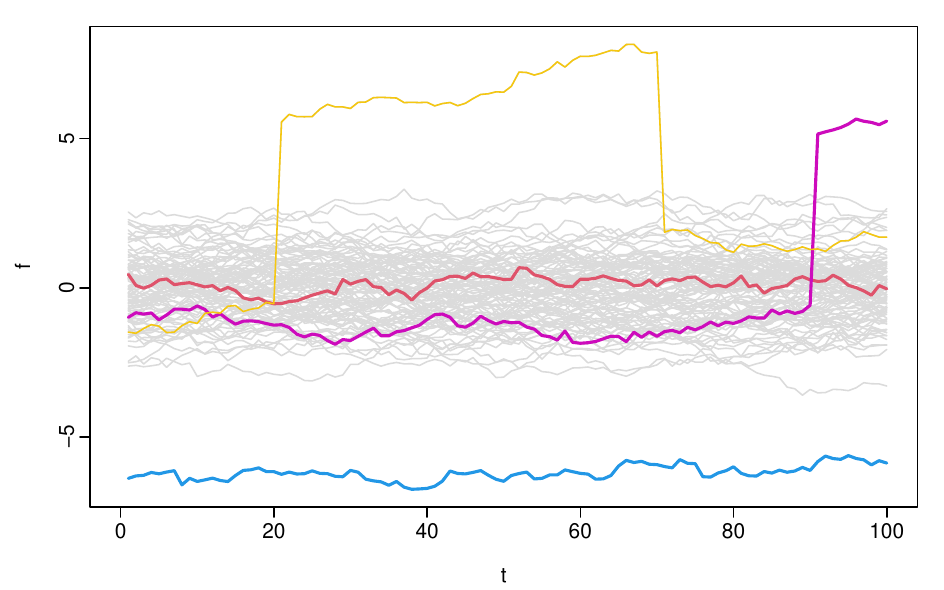}
\qquad
\includegraphics[width=70mm]{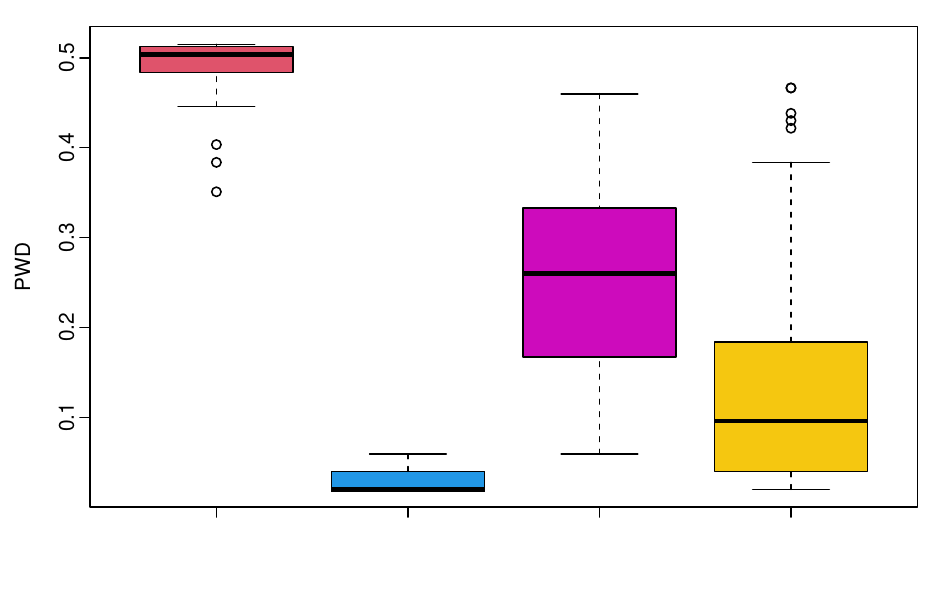}
    \caption{Left panel shows 100 simulated curves with 3 magnitude outliers and the median curve highlighted in colors. The grey curves are the non-outlying functional curves. The right panel shows the boxplot of the univariate $\widehat{\text{PWD}}$ distribution for the median curve as well as for the magnitude outliers.}
    \label{fig1: PWD pure magnitude}
\end{figure}

The efficient and fast computational algorithm proposed by \cite{comp} for computing MBD  directly calculates the $\widehat{\text{PWD}}$ at all $t$. Instead of reducing to one depth value for each functional observation, we propose to use the distribution of the $\widehat{\text{PWD}}$s in time to visualize magnitude outliers. The left panel of Figure  \ref{fig1: PWD pure magnitude} shows 100 simulated curves with 3 magnitude outliers and the median curve highlighted in colors. The grey curves are the non-outlying functional curves, and the red curve is the functional median with the largest MBD value. The right panel of Figure \ref{fig1: PWD pure magnitude} shows the boxplot of the univariate $\widehat{\text{PWD}}$ distribution for the median curve as well as for the magnitude outliers. Regarding the $\widehat{\text{PWD}}$ boxplot, many interesting and detailed features can be explained from this visualization tool, for instance, the magnitude outlyingness, that is, the boxplot of the $\widehat{\text{PWD}}$ for the median curve in red shows the $\widehat{\text{PWD}}$ values are overall high for all time points. Those in magenta and yellow both are partially magnitude outliers due to the large variability among the $\widehat{\text{PWD}}$ in time. The magenta boxplot has a higher median indicating the outlyingness is present for a shorter time period. To visualize magnitude outliers, the distribution of the $\widehat{\text{PWD}}$ of any given curve is compared to such distribution for the median curve. Although formal hypothesis tests can be developed to compare two sets of distributions of the $\widehat{\text{PWD}}$, it is not necessary for visualization purposes.

\noindent
 Functional shape outliers are observations that exhibit temporal patterns distinct from the majority. Intuitively, these patterns can be characterized by their evolution over time.
This underpins the rationale for employing the first-order autocorrelation of the pointwise depth ($\widehat{\text{PWD}}$) to highlight the distinctive features of a function's shape.

By examining the distribution of the $\widehat{\text{PWD}}_i(t_j;S)$ for $i=1,\ldots,n,j=1,\ldots,p$, shape outliers cannot be identified because the boxplot does not consider the time order of these $\widehat{\text{PWD}}_i(t_j;S)$ values. Therefore, we define pairwise depth 
\begin{equation*}
    \text{PD}_{i}(t;S):=\left(\widehat{\text{PWD}}_{i}(t_j;S),\widehat{\text{PWD}}_{i}(t_{j+1};S) \right),~i=1,\cdots,n,j=1,\ldots,p-1,
\end{equation*}
for each function $X_i(t)$  and propose to use the bivariate random vector of $\widehat{\text{PWD}}_i(t_j; S)$, or the scatter-plot with $p-1$ time points, and the sample correlation of the $\text{PD}_{i}(t; S)$ as a measure of shape outlyingness. For simplicity, the sample correlation is denoted as $r_{i}:=\text{cor}\left(\text{PD}_{i}(t;S)\right),i=1,\ldots,n,~j=1,\ldots,p-1$.

One great advantage of the proposed method is that the outlier detection procedure can be easily extended to multivariate functional data, when the $\widehat{\text{PWD}}$ is calculated by a depth notion for multivariate data, such as the spatial depth introduced by \cite{Serfling2002}, and the simplicial depth proposed by \cite{simpl}. 

\subsection{Outlier detection procedure and visualization}\label{outlier detection}

Every outlier detection procedure needs a decision rule. For example, the functional boxplot uses the  $1.5 \times 50\%$ central region rule to flag magnitude outliers. The MS plot uses the distribution of the robust Mahalanobis distance. The MSS uses the classical boxplot of the MSS to flag shape outliers. The outliergram uses the distances to the parabola expression between the MBD and EPI to detect shape outliers. 

 In the functional outlier detection literature, functional boxplot has proven to be effective for detecting magnitude outliers \citep[see, e.g.,][]{Sun2011,Narisetty2016,TVD}. In our numerical studies, we use the functional boxplot for magnitude outlier detection and we propose the boxplot of the $\widehat{\text{PWD}}$ for visualization of detected outliers. Therefore,  the simulation settings established in section~\ref{sec:Simulation}  only focus on shape outlier detection. 

\begin{figure}[!ht]
\centering
\includegraphics[width=45mm]{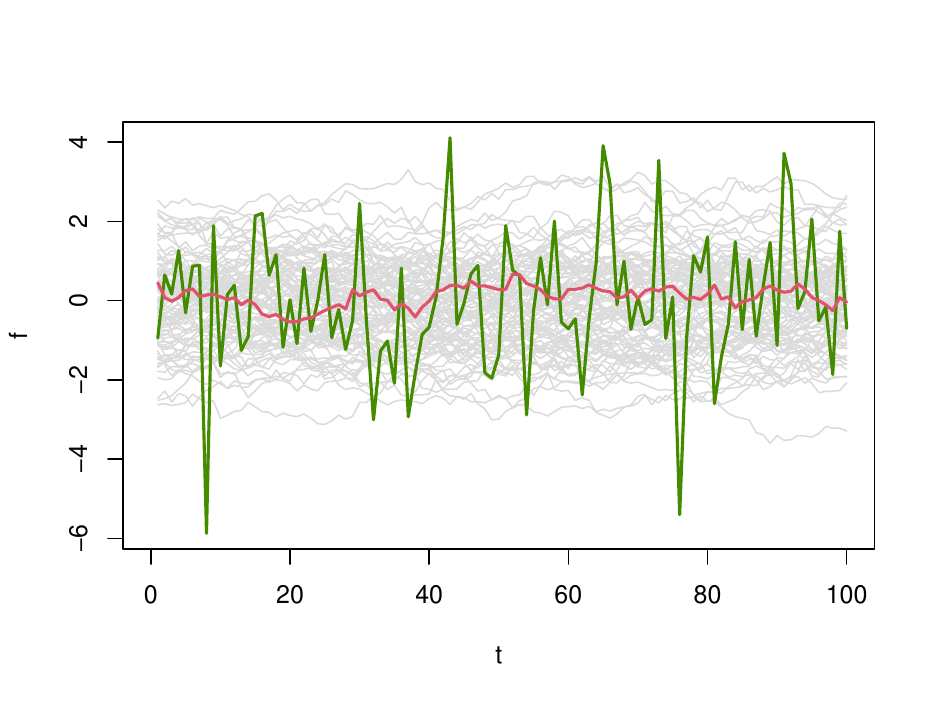}
\qquad
\includegraphics[width=45mm]{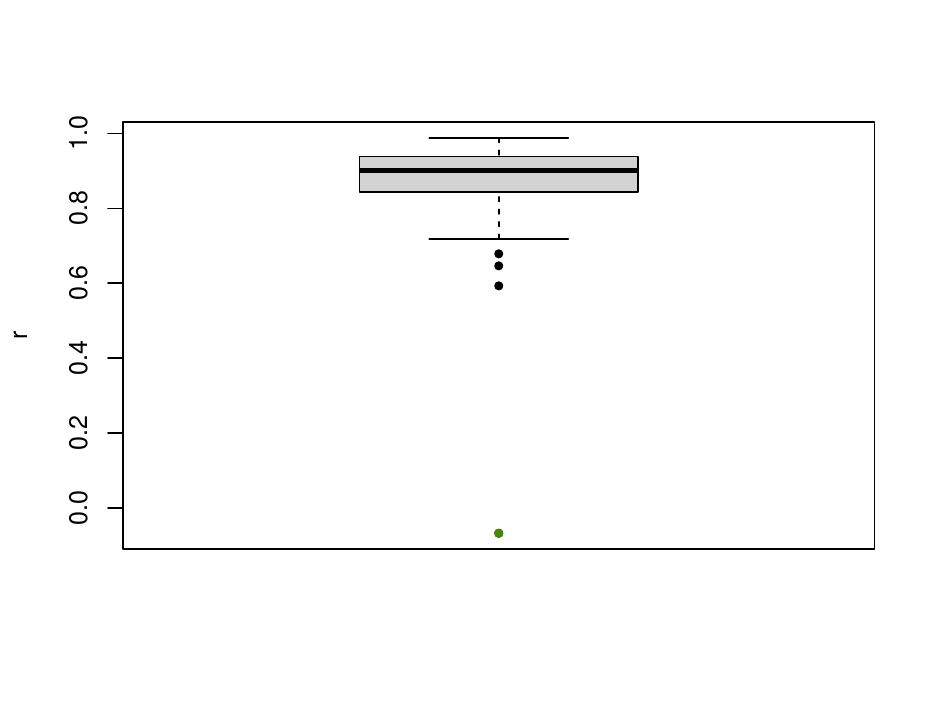}
\qquad
\includegraphics[width=45mm]{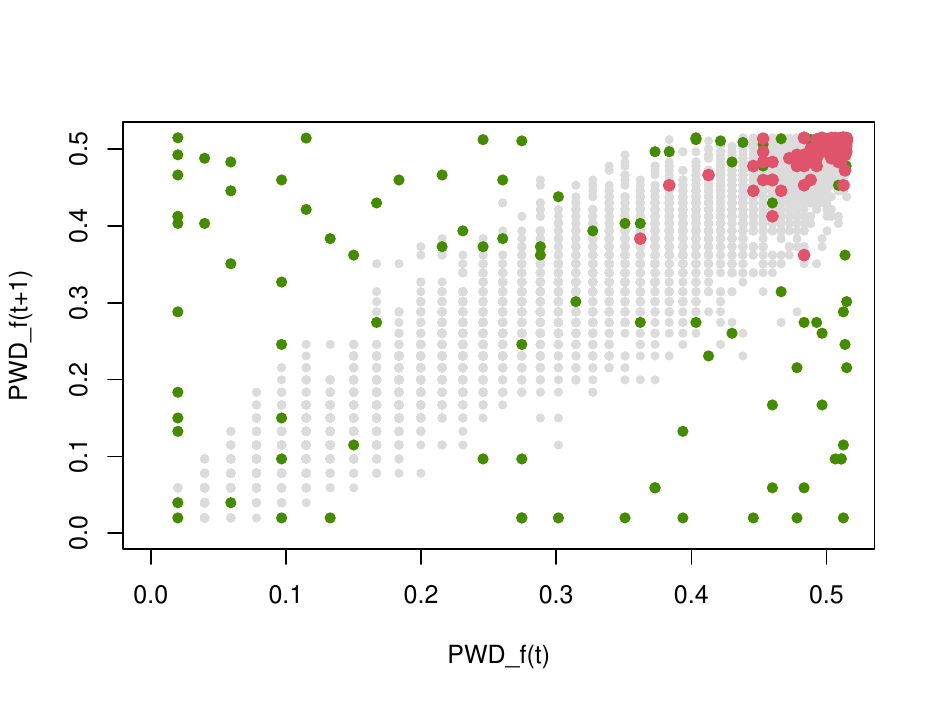}
\\
\includegraphics[width=45mm]{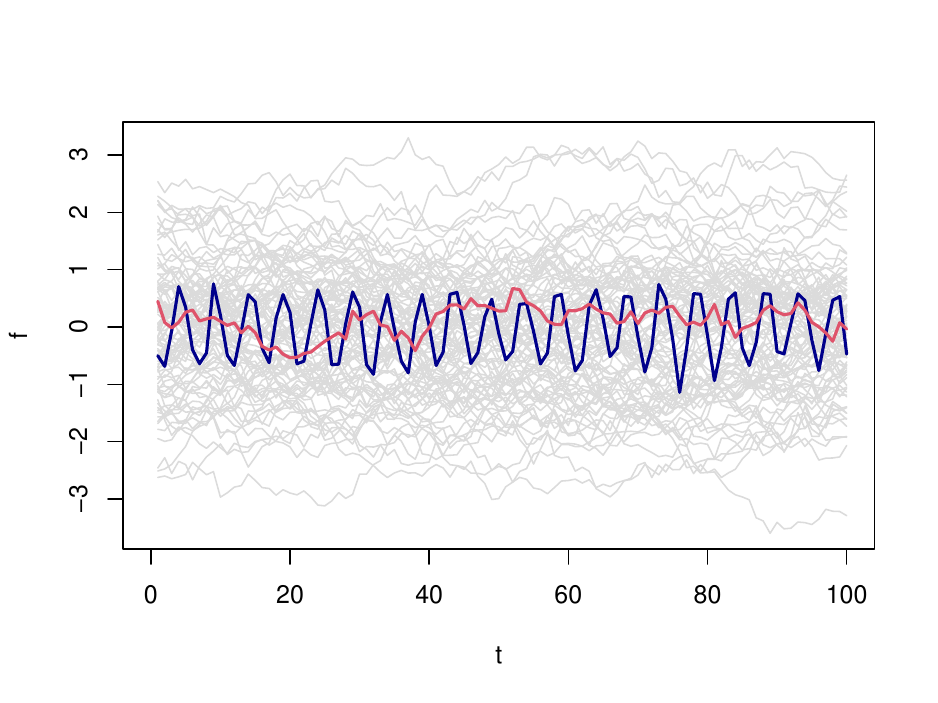}
\qquad
\includegraphics[width=45mm]{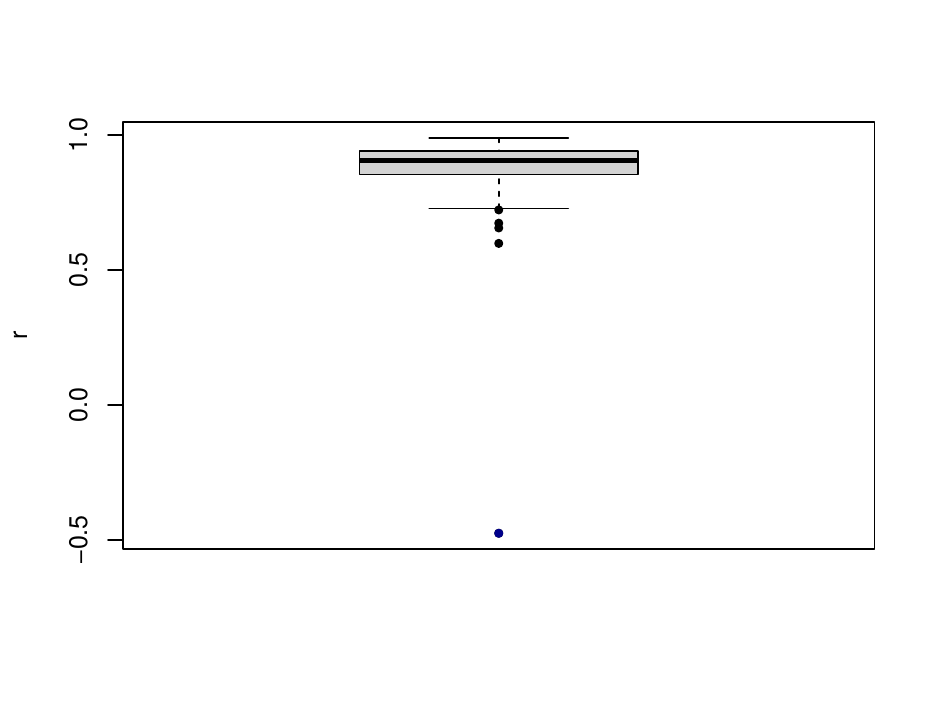}
\qquad
\includegraphics[width=45mm]{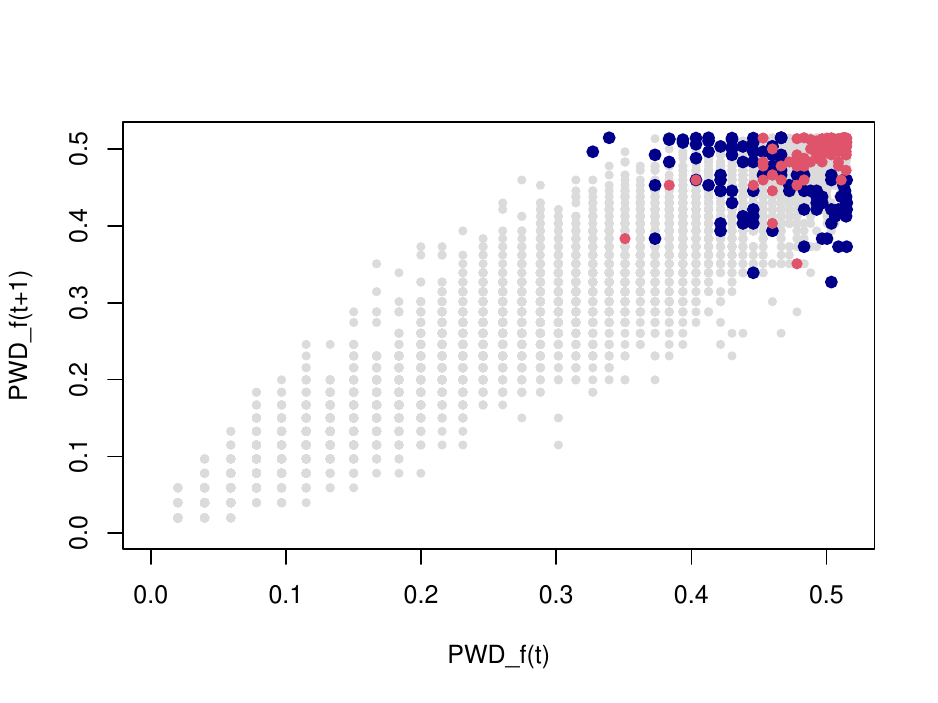}
\\
\includegraphics[width=45mm]{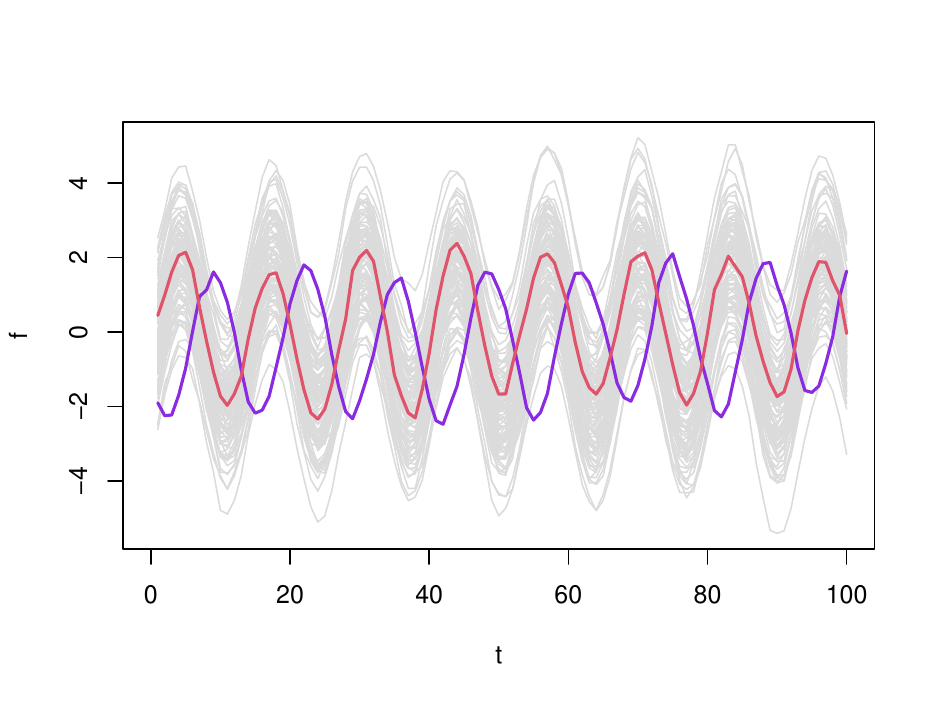}
\qquad
\includegraphics[width=45mm]{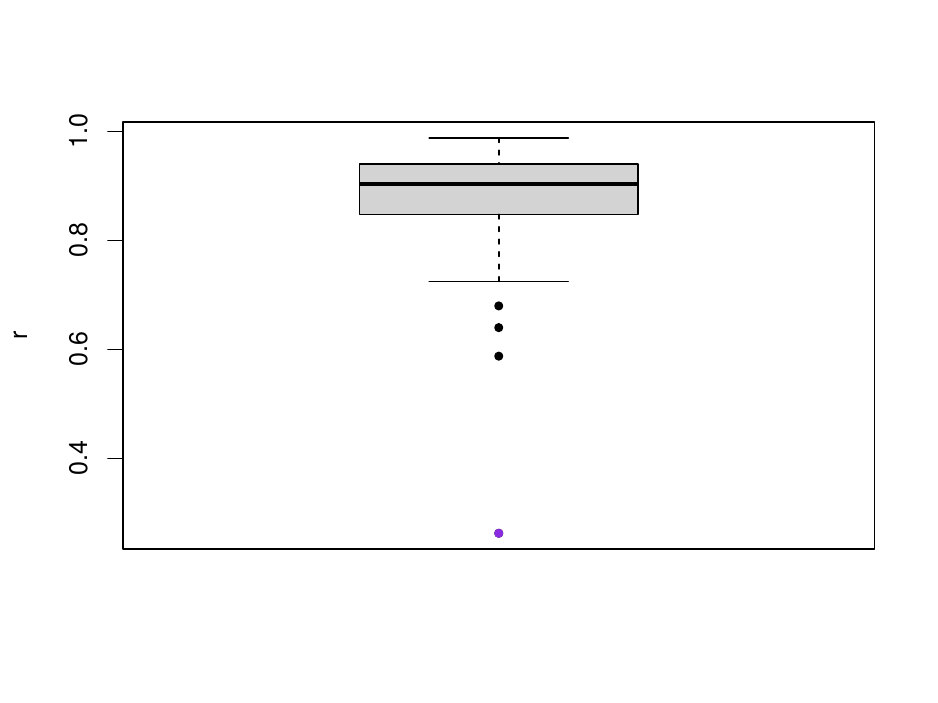} 
\qquad
\includegraphics[width=45mm]{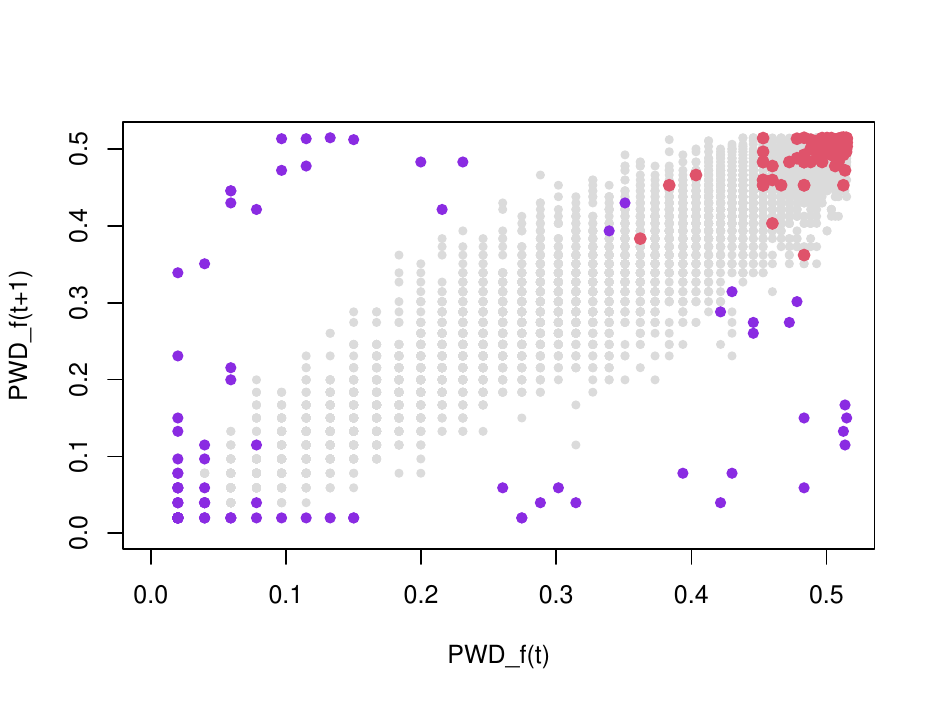}
    \caption{Left panel shows 100 curves for three different shape outlier models. The median curve in all cases is highlighted in red, the shape outlier in each case has a different color. The middle panels represent the boxplot of the sample correlation of the PD for the curves plotted in the left panel. The right panel shows the scatter plot of the bivariate random vector of the  $\widehat{\text{PWD}}$. The median curve is highlighted in red, all the shape outliers are also highlighted in colors.}
    \label{fig2:PWD and PD for shape outlier models}
\end{figure}

We have introduced the sample correlation of the $\text{PD}$ (denoted as $r_i$) in Section~\ref{2.1} to measure the shape outlyingness of a given curve. Regarding shape outlier detection, we propose constructing a boxplot for $r_i$ and using the $F \times \text{IQR}$  empirical rule of the boxplot for outlier detection, where $F$ is the empirical factor. In this paper, all simulations are carried out with  $F = 3$ as in the boxplot for MSS in \cite{TVD}. Curves with the sample correlation exceeding the lower fence of the boxplot are flagged as outliers.  Fig~\ref{fig2:PWD and PD for shape outlier models} shows the $r_i$ boxplots and the $\text{PD}$ scatter plots. Here in each of the left panels, the gray curves correspond to the non-outlying curves, and the red curve is the median curve for each of the models with the maximum value of the MBD. The green curve in the top left panel is an example of a dependence shape outlier (presented as Model 1 in Section~\ref{sec:Simulation}), characterized by abrupt variations over time. The navy curve in the center-left panel depicts a shape outlier characterized by frequent fluctuations around the median curve. Finally, the purple curve in the bottom left panel is a shape outlier, which is distinguished by a shifted version of the non-outlying curves. The middle panels show the corresponding boxplots of the sample correlations of the curves represented in the left panels, where the detected outliers match the outlier curve in the left panel. The scatter plots in the right panels show that the $\text{PD}$ of the non-outlying curves is placed around the $45$-degree line and, likewise for the median curve. However, as expected for the shape outliers, their bivariate distribution is inconsistent, and their values are farther from the $45$-degree line in the scatter plot. 
 In the case of multivariate functional data, we use an identical approach to identify outliers by using the sample correlation of the bivariate random vector of the $\widehat{\text{PWD}}_i(t;S)$, the main difference being the concept of multivariate depth to be applied. In particular, in section~\ref{Sim:multivariate}, we recommend applying the concept of simplicial depth from \cite{simpl}. 

\subsection{Existence of shape outliers: a bootstrap testing procedure} \label{meth:overall}

Many existing methods, such as the outliergram, the MS plot, and the MSS, tend to have a high false positive rate when no shape outliers exist. We propose a bootstrap-based test of the hypothesis procedure to determine whether at a given significance level $\alpha$ exists any shape outlier. This test can be done before applying any shape outlier detection methods. Let $S:=\{X_1(t),X_2(t),\ldots,X_n(t)\}$ be a set of $n$ random realizations of a stochastic process $X$.  Let $\{\text{PD}_1(t),\text{PD}_2(t),\ldots,\text{PD}_n(t)\}$ and $\{r_1,r_2,\ldots,r_n\}$ be the pairwise depth (PD) and the sample correlation ($r$) of the random realizations $\{X_1(t), X_2(t),\ldots, X_n(t)\}$, respectively. The computation of such quantities is described in Section \ref{2.1}. We want to test for the null hypothesis: 
 \begin{center}
   $H_0$:  ~There are no shape outliers in $S$\\
   $H_1$:  ~There is at least one shape outlier in $S$.\\ 
\end{center}
The testing procedure is sequential and should be carried out as follows:

\begin{enumerate}
 \item \label{TS} \textbf{Test statistic}. Let $\mathbb{T}$ be the overall test statistics among all the $n$ realizations in $S$ of $X$. $\mathbb{T}$ is defined as follows:
 \begin{equation*}
     \mathbb{T}:=\frac{|\min{(r_{i})}-1|}{\text{sd}(r_{i})},~\text{for}~i=1,\ldots,n,
 \end{equation*}
where, $\min{(r_{i})}:=\min(r_1,\ldots,r_n)$ denotes the minimum value of the sample correlation across the $n$ random realizations. $\text{sd}(r_{i}):=\text{sd}(r_1,\ldots,r_n)$ stands for the standard deviation across all the $n$ sample correlations. 

\item \textbf{Reference distribution under $\mathbf{H_0}$}. In order to test for the null hypothesis $H_0$, we need to determine or estimate a reference distribution of $\mathbb{T}$ under $H_0$. To this end, let $ \mathds{F}$ be the distribution of $\mathbb{T}$ under $H_0$.  Because $\mathds{F}$ may not be a known distribution, $\mathds{F}$ may need to be constructed. It is established that under $H_0$, all realizations result from a stochastic process with no shape outliers. To this end, we proposed a bootstrap-based method to estimate a distribution function $\widehat{\mathds{F}}$. The bootstrap procedure to obtain $\widehat{\mathds{F}}$ is as follows:
\begin{enumerate}[a)]
    \item The first stage is building a robust stochastic process whose realizations cannot contain shape outliers.  We propose using a Gaussian stochastic process whose parameters can be robustly estimated from the realizations of $X$. Let $Z(t)$ be a Gaussian stochastic process with a mean and covariance matrix defined as $\bm{\mu}$ and $\bm{\Sigma}$ respectively. i.e., $Z(t)\sim \mathcal{N}(\bm{\mu},\bm{\Sigma})$.

    \item To be able to simulate realizations from $Z(t)$, we need to fully characterize its distribution, i.e., we need to robustly estimate $\bm{\mu}$ and $\bm{\Sigma}$ respectively. That is, let $\widehat{\bm{\mu}}$ and $\widehat{\bm{\Sigma}}$  denote the robust estimators of the vector mean $\bm{\mu}$ and covariance matrix $\bm{\Sigma}$ of the Gaussian process $Z$. Thus, with $X_1(t),X_2(t),\ldots,X_n(t)$, we propose estimate $\widehat{\bm{\mu}}$ and  $\widehat{\bm{\Sigma}}$ as follows:
    
    \begin{enumerate}
        \item The robust estimator  $\widehat{\bm{\mu}}$ based on the $n$ random realizations from $X$ can be achieved by estimating the pointwise median.
        \item \label{Qn} In order to estimate $\bm{\Sigma}$, and since outliers may exist in $S$, robust techniques are required. To robustly estimate the covariance matrix $\bm{\Sigma}$, we propose the $\mathcal{Q}_n$ componentwise estimator of a dispersion matrix by \cite{Genton2001}. This estimator is location-free, and previous research has successfully used it in different areas. In spatial statistics with variogram estimation \citep{Genton1999}, and in time series analysis with autocovariance function estimation \citep{Genton2000}.
         \item Fit a parametric covariance model (e.g., exponential, Mat\'ern, rational quadratic, etc.) to those $\mathcal{Q}_n$ obtained in the step \ref{Qn}. 
    \end{enumerate}

    \item Simulate a set of $B$ bootstrap samples from $\widehat{Z}(t) \sim \mathcal{N}(\widehat{\bm{\mu}},\widehat{\bm{\Sigma}})$, i.e., $\{\widehat{Z}_1^{*}(t),\widehat{Z}_2^{*}(t),\ldots,\widehat{Z}_B^{*}(t)\}$. For each of the bootstrap samples $\widehat{Z}_i^{*}(t),i=1,\ldots, B$, compute the test statistics defined in step \ref{TS}. Finally, construct the empirical distribution function $\widehat{\mathds{F}}$ based on $\mathbb{T}_i,i=1,\ldots,B$.

\end{enumerate}

 \item \textbf{Critical value from the reference distribution under $\mathbf{H_0}$}. Based on the empirical distribution function $\widehat{\mathds{F}}$ of $\mathbb{T}$, the critical value is chosen to be the $100(1-\alpha)\%$ quantile. $\alpha$ is defined as the probability of the type I error.
 
\item \textbf{Decision rule}. Reject $H_0$ if $\mathbb{T} \geq \widehat{\mathds{F}} (1-\alpha)$.

\end{enumerate}

This procedure of robustly estimating the covariance matrix in \ref{Qn} is similar to the one in \cite{AdjSun2012}  when selecting the adjustment factors for the adjusted functional boxplot. The reason for fitting a parametric covariance model to the estimated covariances is to perform a parametric bootstrapping procedure to determine the critical value. We will use simulations for the assessment of the Type I error in section \ref{sub:typeI}.

\section{Simulation Study}\label{sec:Simulation}
In this section, we compare the performance of the proposed method with several functional data models for both univariate and multivariate cases through an extensive simulation study. The comparison is carried out with the most recent functional outlier detection methods. 

\subsection{Univariate functional data}

For the univariate case,  most outlier models introduced here are originated from \cite{MBD,Sun2011,outliergram,Daitransf}; however, we made some modifications and introduced other outlier models. The five models refer to shape outliers.  Here in all simulations $\mathcal{T}=[0,1]$. For the simulation study presented in this section, the contamination rate, or the proportion of simulated shape outliers in each model is defined as $\theta$.

 \begin{enumerate}[Model 1:]
 
     \item \textbf{Dependence shape outliers}: Some of the realizations are generated from a Gaussian process with a different covariance function. \\
      $X_i(t)=(1-c_i)e_i(t)+c_i\Tilde{e}_i(t)$, for $i=1,\cdots,n$, where $e_i(t)$ is a zero-mean Gaussian process with a covariance function $c(s,t)=\exp\{-|s-t| \}$, and $\Tilde{e}_i(t)$ is another zero-mean Gaussian process but with a different covariance function $\Tilde{c}(s,t)=6\exp\{-|s-t|^{0.1} \}$, $c_i\sim \text{Bernoulli}(\theta)$. 
      
     \textbf{Remark:} The dependence shape outliers are defined here as a stochastic process whose covariance function presents an increase in the marginal variance from $1$ to $6$ and a reduction in the distance power from $1$ to $0.1$. We include an additional simulation study in the Appendix~ \ref{Appendix_1}, to assess the sensitivity of the proposed approach to this type of shape outliers.

     \item \textbf{Phase shape  outliers}: some realizations are phase deviated from the normal curves.\\ 
      $X_i(t)=(1-c_i)[2\sin(15\pi t)]+c_i[2\sin(15 \pi t +4)]+e_i(t)$, for $i=1,\cdots,n$,
      where $c_i$ and $e_i(t)$ have the same definitions as before.
     
     \item \textbf{High frequency low amplitude}: some realizations have very low amplitude with very high frequency. \\
      $X_i(t)=(1-c_i)[0.1+\arctan(t)+e_i(t)]+c_i[ \arctan(t)+\epsilon^*_i(t)]$, for $i=1,\cdots,n$,
      where $\epsilon^*_i(t)$ is another zero-mean Gaussian process with a different covariance function $c^*(s,t)=0.1\exp\{-\frac{|s-t|^{0.1}}{4}\}$, $c_i$ and $e_i(t)$ have the same definitions as before.
      
     \item \textbf{Slightly modification to the outlier model proposed by \cite{outliergram}}: some realizations have very low amplitude with very high frequency. \\
      $X_i(t)=(1-c_i)[30t(1-t)^{3/2}+e_i(t)]+c_i[30t(1-t)^{3/2}+\epsilon^*_i(t)]$, for $i=1,\cdots,n$
      where $\epsilon^*_i(t),c_i$ and $e_i(t)$ have the same definitions as before.
    
     \item \textbf{Central high frequency-low amplitude}: some realizations have frequent changes around the median. \\
    $X_i(t)=(1-c_i)[e_i(t)]+c_i[0.1\sin(40(t+\Theta)\pi)+\epsilon^*_i(t)]$, for $i=1,\cdots,n$, where
    
   $\Theta \sim \text{Uniform}[0.25,0.5]$. 
     $\epsilon^*_i(t),c_i$ and $e_i(t)$ have the same definitions as before.
 \end{enumerate}
 
In all the simulation scenarios, we have set the contamination rate $\theta$ to a fixed value of $\theta=0.1$. In the appendix~\ref{Appendix_2}, we have conducted a detailed analysis of the sensitivity of each of the previously discussed functional outlier models to an increasing trend in the contamination rate.

We compare the performance of our proposed outlier detection rule for univariate functional data, with other outlier detection rules in the literature. We consider, the following methods:
\begin{itemize}
    \item MS plot: is used to detect univariate functional outliers \citep{msplot}.
    \item TVD and MSS:  shape outliers are detected by the classical boxplot of the MSS \citep{TVD}.
    \item OG: shape outliers detected through the outliergram defined by \cite{outliergram}.
    \item MBD: the modified band depth defined in \cite{MBD}, is used in the functional boxplot to detect outliers. 
    \item ED: the extremal depth defined in \cite{Narisetty2016} is used in the functional boxplot to detect outliers. 
\end{itemize}

Table \ref{Table1-Univ_results} presents the simulation results of our proposed method for functional outlier detection, in comparison with other existing methods. The evaluation was performed with a contamination rate set to $\theta=0.1$. The performance is assessed by the mean and standard deviation of the true and false positive rates. The true positive rate (TPR) is defined as the ratio of the number of correctly detected outliers to the number of true outliers present in the simulation. The false positive rate (FPR) is defined as the ratio of incorrectly detected outliers to the number of true non-outlier curves. Both $\text{TPR}$ and $\text{FPR}$ are defined in Eq. \eqref{TPR-FPR}.

\begin{equation}
    \begin{aligned}
    \text{TPR}_i&=\frac{ \text{Number of true detected outliers}}{ \text{Number of true outliers}}\\
    \text{FPR}_i&=\frac{ \text{Number of false detected outliers}}{\text{ Number of Non-outlying curves}},~i=1,\ldots,500.
    \label{TPR-FPR}
\end{aligned}
\end{equation}

 The simulation results reported in Table \ref{Table1-Univ_results} are based on $500$ simulations for each model, results include the means and standard deviations (in parenthesis) shown as percentages. 

We can observe that the proposed method is outstanding compared to all methods, especially for models 3-5, where there are shape outliers with low amplitude fluctuations around the functional median. For such models, all our competitors show a very low value of the mean TPR, since they cannot detect this kind of outliers. 
For Models 1 and 2, both our method and MS plot perform well. 

\begin{table}[!ht]
    \centering
     \caption{Results of outlier detection using different methods for different models. The values are the TPR and FPR means shown as percentages in the 500 experiments for each case, and the values in parentheses are the corresponding standard deviations.}
     \resizebox{\textwidth}{!}{
    \begin{tabular}{@{}lllllllllll@{}}
    \toprule
          & \multicolumn{2}{c}{\textbf{Model 1}} & \multicolumn{2}{c}{\textbf{Model 2}} & \multicolumn{2}{c}{\textbf{Model 3}} & \multicolumn{2}{c}{\textbf{Model 4}} & \multicolumn{2}{c}{\textbf{Model 5}} \\ \cmidrule{2-11}
     \textbf{Method}  & \textbf{TPR} & \textbf{FPR} & \textbf{TPR} & \textbf{FPR} & \textbf{TPR} & \textbf{FPR} & \textbf{TPR} & \textbf{FPR} & \textbf{TPR} & \textbf{FPR}  \\ \midrule
     \textbf{Proposed} & 100.00 (0.00) & 2.81 (1.54) & 99.28 (2.98) & 2.40 (1.41) & 99.54 (2.34) & 2.67 (1.50) & 99.59 (2.07) & 2.64 (1.51) & 99.89 (1.08) & 2.67 (1.48) \\
     \textbf{MS plot} & 100.00 (0.00) & 2.29 (1.98) & 100.00 (0.00) & 2.22 (1.86) & 0.00 (0.00) & 2.12 (1.83) & 0.00 (0.00) & 2.07 (1.83) & 0.00 (0.00) & 2.24 (1.93) \\
     \textbf{TVD} & 81.27 (14.55) & 0.05 (0.24) & 70.86 (21.92) & 0.05 (0.24) & 0.12 (1.27) & 0.33 (0.71) & 0.00 (0.00) & 0.31 (0.70) & 0.24 (1.94) & 0.30 (0.68) \\
     \textbf{OG} & 75.13 (14.79) & 2.05 (1.68) & 99.87 (1.02) & 1.85 (1.64) & 0.00 (0.00) & 4.29 (2.36) & 0.00 (0.00) & 4.29 (2.32) & 0.00 (0.00) & 4.44 (2.34) \\
     \textbf{MBD} & 79.00 (16.27) & 0.05 (0.25) & 68.26 (24.03) & 0.05 (0.23) & 0.00 (0.00) & 0.33 (0.71) & 0.00 (0.00) & 0.33 (0.71) & 0.00 (0.00) & 0.32 (0.68) \\
     \textbf{ED} & 76.10 (16.30) & 0.02 (0.16) & 55.01 (25.32) & 0.02 (0.15) & 0.00 (0.00) & 0.15 (0.42) & 0.00 (0.00) & 0.15 (0.42) & 0.00 (0.00) & 0.15 (0.42) \\
     \bottomrule
    \end{tabular}}
    \label{Table1-Univ_results}
\end{table}

\subsection {Multivariate functional data} \label{Sim:multivariate}

We design three simulation scenarios for functional outlier detection for multivariate functional data. The outlier models introduced in this section originated from \citep{Dai2018,msplot}; however, we made some modifications to those outlier models. We study three models in total, and such models are high-frequency shape outliers. We set a contamination rate $\theta=0.1$. Here in all simulations $\mathcal{T}=[0,1]$.
 The Non-outlying observations $X_i(t):=(X_{i1}(t),X_{i2}(t))^\top=e_i(t)$, where $e_i(t)=(e_{i1}(t),e_{i2}(t))^\top~i=1,\ldots,n$, is defined as  a bivariate  Gaussian process with zero mean and cross-covariance function proposed by~\cite{matern}:
     \begin{equation}
         c_{kl}(s,t)=\rho_{kl}\sigma_k\sigma_l\mathcal{M}(|s-t|;\nu_{kl},\gamma_{kl}),~~k,l=1,2,
     \end{equation}
\noindent where $\rho_{12}$ is the correlation between $X_{i1}(t)$ and $X_{i2}(t)$, $\rho_{11}=\rho_{22}=1$, $\sigma_k^2$ is the marginal variance and $\mathcal{M}(h;\nu,\alpha)=2^{1-\nu}\Gamma(\nu)^{-1}(\gamma|h|)\mathcal{K}_{\nu}(\gamma|h|)$ with $|h|=|s-t|$, is the Mat\'ern class introduced by \cite{matern2013spatial}, $\mathcal{K}_{\nu}$ is the modified Bessel function of second kind, $\nu>0$ is a smoothness parameter, and $\gamma>0$ is a range parameter. Similar to \cite{Dai2018,msplot}, we set $\sigma_1=\sigma_2=1,\gamma_{11}=0.02,\gamma_{22}=0.01,\gamma_{12}=0.016,\nu_{11}=1.2,\nu_{22}=0.6,\nu_{12}=1$ and $\rho_{12}=0.6$.  The outlier models are defined as follows 

  \begin{enumerate}[Model 1:]\label{Multi:nonout} 
    \item \textbf{High frequency low amplitude}: some of the realizations have frequent changes around the median with low amplitude. For each outlier function, each component has a different frequency.\\
    $X_i(t):=(X_{i1}(t),X_{i2}(t))^\top=(1-c_i)e_i(t)+c_i[(0.5\cos(80\pi t)+e_{i1}(t),0.75\sin(40\pi t)+e_{i2}(t))^\top]$, for 
    $i=1,\ldots,n$, where $c_i\sim \text{Bernoulli}(\theta)$. 
    
    \item \textbf{High frequency high amplitude}: some of the realizations have frequent changes around the median with high amplitude. For each outlier function, each component has a different frequency.\\

    $X_i(t):=(X_{i1}(t),X_{i2}(t))^\top=(1-c_i)e_i(t)+c_i[(2\cos(80\pi t)+e_{i1}(t),3 \sin(40\pi t)+e_{i2}(t))^\top]$, for 
    $i=1,\ldots,n$, where $c_i$ has the same definition as before.

    \item \textbf{Same high frequency for both components}: some of the realizations have frequent changes around the median with low amplitude. For each outlier function, each component has the same high frequency.\\

    $X_i(t):=(X_{i1}(t),X_{i2}(t))^\top=(1-c_i)e_i(t)+c_i[(\cos(40\pi t)+e_{i1}(t), \sin(40\pi t)+e_{i2}(t))^\top]$, for 
    $i=1,\ldots,n$, where $c_i$ has the same definition as before.

 \end{enumerate}

For multivariate functional outlier detection, we consider one possible competitor to the proposed method. The magnitude-shape plot (MS plot) proposed by \cite{Dai2018}, which in turn is constructed based on the notion of functional directional outlyingness introduced by \cite{msplot}. Table 2 shows the simulation results for the three models we considered in the multivariate case. We can observe that our method outperforms the MS plot in detecting this type of shape outlier. For models 1 and 3, our method always detects all shape outliers with a TPR of $100\%$.

\begin{table}[!ht]
\caption{Simulation results of outlier detection using the proposed and MS plot methods for different models. The values are the means shown as percentages in the 500 experiments for each case, and the values in parentheses are the corresponding standard deviations.}
\centering
\resizebox{\columnwidth}{!}{
\begin{tabular}{|c|cc|cc|cc|ll}
\cline{1-7}
\multirow{2}{*}{\textbf{Method}} & \multicolumn{2}{c|}{\textbf{Model 1}}            & \multicolumn{2}{c|}{\textbf{Model 2}}             & \multicolumn{2}{c|}{\textbf{Model 3}}             & \multicolumn{2}{c}{\textbf{}} \\ \cline{2-7}
                                 & \multicolumn{1}{c|}{\textbf{TPR}} & \textbf{FPR} & \multicolumn{1}{c|}{\textbf{TPR}}  & \textbf{FPR} & \multicolumn{1}{c|}{\textbf{TPR}}  & \textbf{FPR} & \textbf{}          &          \\ \cline{1-7}
\textbf{Proposed}                & \multicolumn{1}{c|}{99.90 (0.98)} & 1.51 (0.81)  & \multicolumn{1}{c|}{100.00 (0.00)} & 1.54 (0.78)  & \multicolumn{1}{c|}{100.00 (0.00)} & 1.49 (0.72)  &                    &          \\ \cline{1-7}
\textbf{MS plot}                 & \multicolumn{1}{c|}{0.56 (2.48)}  & 0.19 (0.50)  & \multicolumn{1}{c|}{100.00 (0.00)} & 0.22 (0.54)  & \multicolumn{1}{c|}{1.33 (4.21)}   & 0.13 (0.44)  &                    &          \\ \cline{1-7}
\end{tabular}
}
\end{table}

\subsection{Type I Error}\label{sub:typeI}

In Section~\ref{meth:overall}, we describe a bootstrap-based testing procedure for testing the existence of shape outliers. We conducted a simulation study to verify the Type I error of the proposed testing procedure. Table~\ref{table3:typeI_results} shows the simulation results for these different simulation scenarios. Let $n$ be the total number of non-outlying curves and $\alpha$ the probability of rejecting the null hypothesis when it is true.

\begin{table}[!ht]
\caption{Simulation results to the Type I error}
\centering
\begin{tabular}{|l|l|l|l|}
\hline
$n$/$\alpha$ & $\alpha=$ 0.01   & $\alpha=$ 0.05   & $\alpha=$ 0.10   \\ \hline
$n=1000$                  & 0.010  &  0.0490 & 0.1010 \\ \hline
$n=10000$                 & 0.0140 & 0.0539  & 0.1071 \\ \hline
$n=100000$                & 0.0128 & 0.0566  & 0.1106 \\ \hline
\end{tabular}
\label{table3:typeI_results}
\end{table}

Based on the results in Table \ref{table3:typeI_results}, we observe that at different $\alpha$ levels and different sample sizes $n$, the testing procedure keeps the nominal $\alpha$ well.

\section{Application: outlier detection in a photovoltaic system}\label{sec:application}

This section applies the proposed outlier detection method and visualization tools to input-output data from a simulated PV system. The simulated data is particularly valuable in PV systems design, enabling performance assessment and estimating PV installation potential in specific locations. Several specialized tools have been developed for evaluating and estimating PV power production, such as Photovoltaic Geographical Information System (PVGIS), PVWatts, and RETScreen~\citep{psomopoulos2015comparative}. For this application, we adopt the PVGIS provided by the European Commission.

\begin{figure}[h!]
  \centering
  \includegraphics[width=12cm]{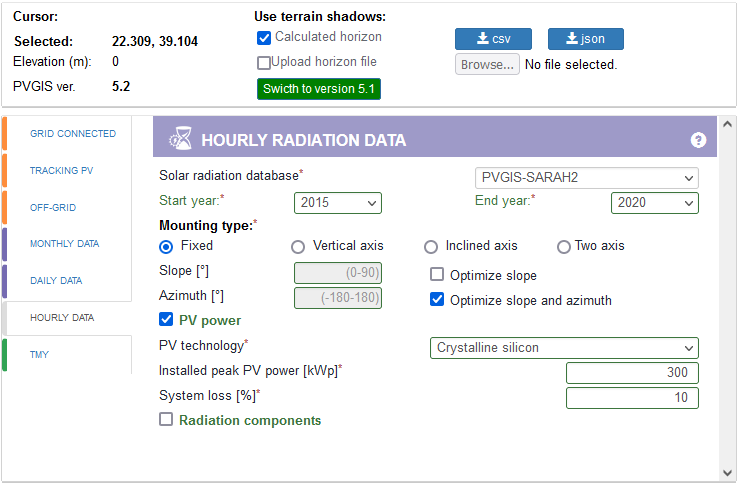} 
  \caption{ PV system simulation via the PVGIS web-interface.}\label{PVGIS}
\end{figure}

The PVGIS web interface, as depicted in Figure~\ref{PVGIS}, is a user-friendly online tool provided by the European Commission. It covers Europe, Africa, most of Asia, and parts of South America, facilitating the assessment and estimation of PV solar energy potential across various regions. This comprehensive platform allows users to access solar radiation and energy output data, aiding in evaluating and planning PV installations. The interface offers features including geographical mapping, energy yield calculations, and performance assessments. With its intuitive design and interactive capabilities, users can easily input parameters such as location coordinates, PV system specifications, and time periods to obtain valuable insights into PV energy production. By leveraging satellite data and sophisticated algorithms, the PVGIS web interface empowers users to make informed decisions regarding PV system deployment, contributing to sustainable energy utilization. Further details about PVGIS's basic features can be found in~\cite{psomopoulos2015comparative,vsuri2005pv}.

This application specifically utilizes PVGIS to estimate the power production of a 300 kWp PV grid-connected system at the King Abdullah University of Science and Technology (KAUST), Saudi Arabia. The simulated PV system is of the fixed-mounted (non-tracking) type, utilizing PV modules based on crystalline silicon technology. As the input data for the PV system, we rely on the PVGIS-SARAH2 solar radiation database, calculated from satellite images. This database, available in PVGIS, provides hourly data and covers the period from 1 January 2005 to 31 December 2020~\citep{araveti2022wind}.


\medskip

\begin{figure}[h!]
  \centering
  \includegraphics[width=7cm]{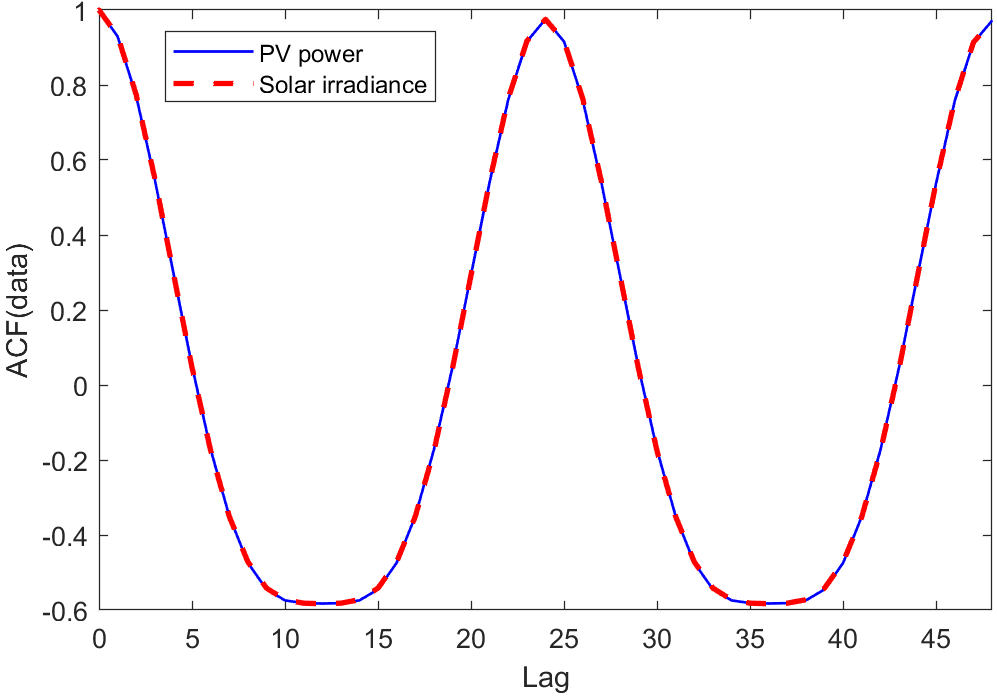} 
  \caption{Sample ACF of PV solar power [W] and Global solar irradiance $[W/m^2]$.}
  \label{ACFData}
\end{figure}

The sample autocorrelation function (ACF) of the data collected from January 1, 2015, to December 31, 2020, is depicted in Figure \ref{ACFData}. We observe the presence of daily periodicity in the data, likely attributed to the diurnal solar irradiance cycle. Additionally, a high degree of similarity between the two variables is evident. This similarity is a consequence of the strong correlation between solar irradiance and the power produced by a PV system, as PV panels directly convert sunlight into electricity.

\medskip
The Pearson correlation between the two variables is 0.9996, indicating a high positive correlation between PV power and solar irradiance. Importantly, these two variables are crucial in designing and evaluating a PV system as the role of these systems is to convert the energy of sunlight into electric energy. 


\medskip
Outlier or anomaly detection is essential in monitoring PV systems, ensuring their safety and sustained productivity while minimizing power losses. These outliers could be caused by different factors~\citep{alam2015comprehensive}, including the aging or degradation of components like pyranometers used for solar irradiance data collection, the accumulation of dust on PV modules, as well as electrical issues like short-circuits, open-circuits, and partial shading~\citep{alam2015comprehensive}. For instance, dust accumulation on PV modules can reduce their efficiency by blocking sunlight. Consequently, outlier detection techniques can help identify when cleaning or maintenance is needed to optimize energy production. Additionally, partial shading events, whether caused by vegetation growth or other obstructions, can result in irregular performance patterns that should be flagged. As PV systems age, component efficiency may decline, resulting in deviations from anticipated performance levels. For example, a deteriorated solar panel may generate less electricity than its expected capacity. Moreover, anomalies detected in data collected from PV systems can also be attributed to sensor and monitoring equipment deficiencies. Notably, inaccuracies in solar irradiance data may arise from issues with the pyranometer. Thus, integrating outlier detection into a predictive maintenance strategy for PV systems offers significant benefits. By continuously monitoring for anomalies, maintenance activities can be scheduled proactively based on the detected issues, helping to avoid unexpected breakdowns and costly repairs.

In this section, we consider curves as functional data for both PV power and global solar irradiance; such curves are only represented during the daytime, which has non-zero values. We consider daily curves from 15 January to 15 June in 2019 and 2020, respectively. The total amount of considered curves per year is 152. From both  Figures \ref{app:winter2019} and \ref{app:winter2020}, we observe a bell-shaped profile for the two variables. We can observe there is a  high PV generation around noon during the study period.

\subsection{Univariate}\label{app:univ}

\begin{figure}[H]
    \centering
 \includegraphics[width=75mm]{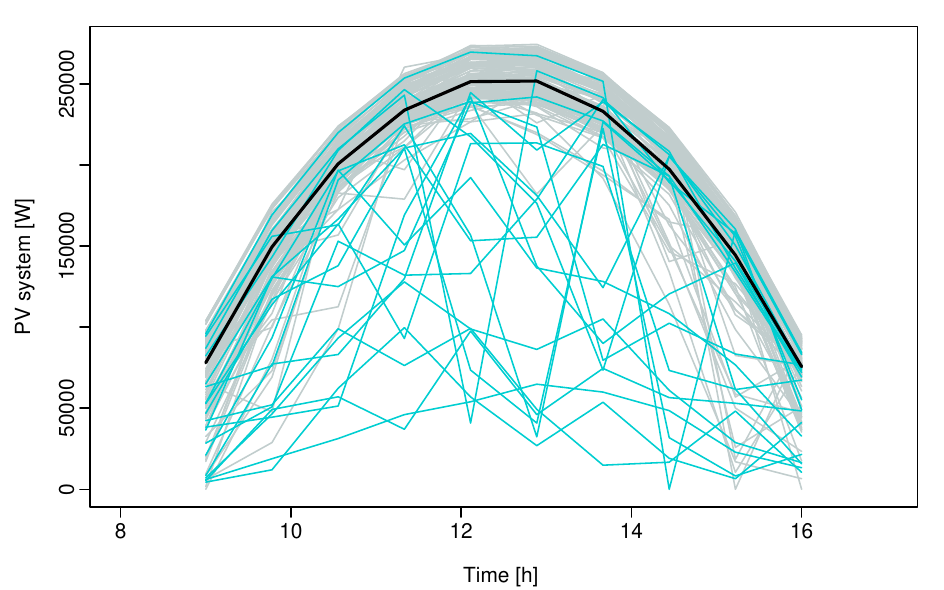}
\quad
  \includegraphics[width=75mm]{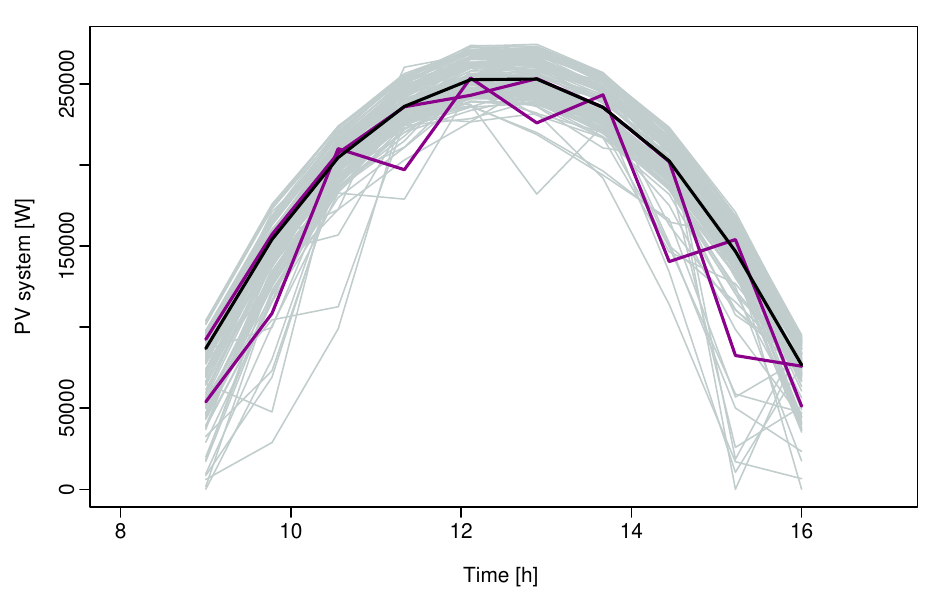}
\hspace{0mm}
  \includegraphics[width=75mm]{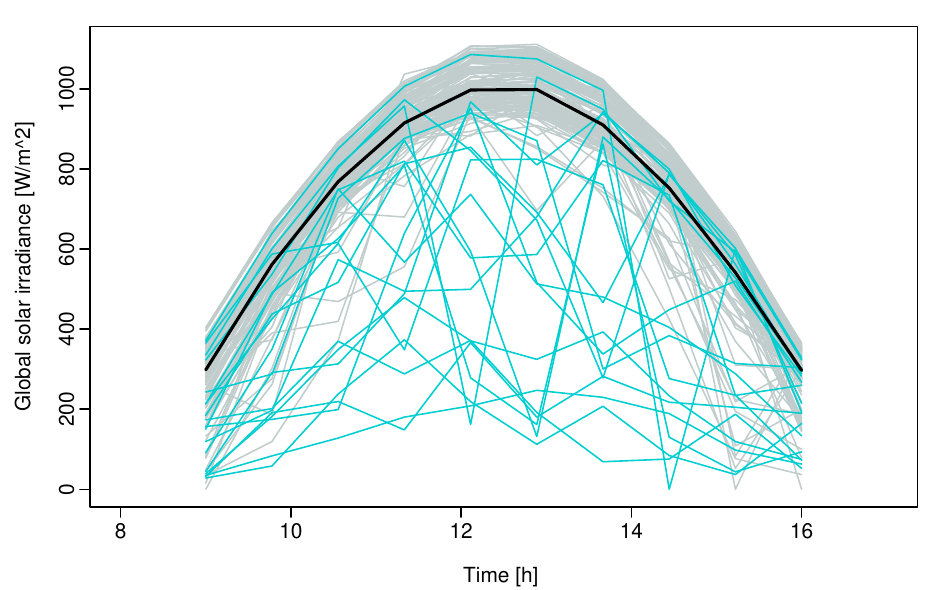}
  \quad
  \includegraphics[width=75mm]{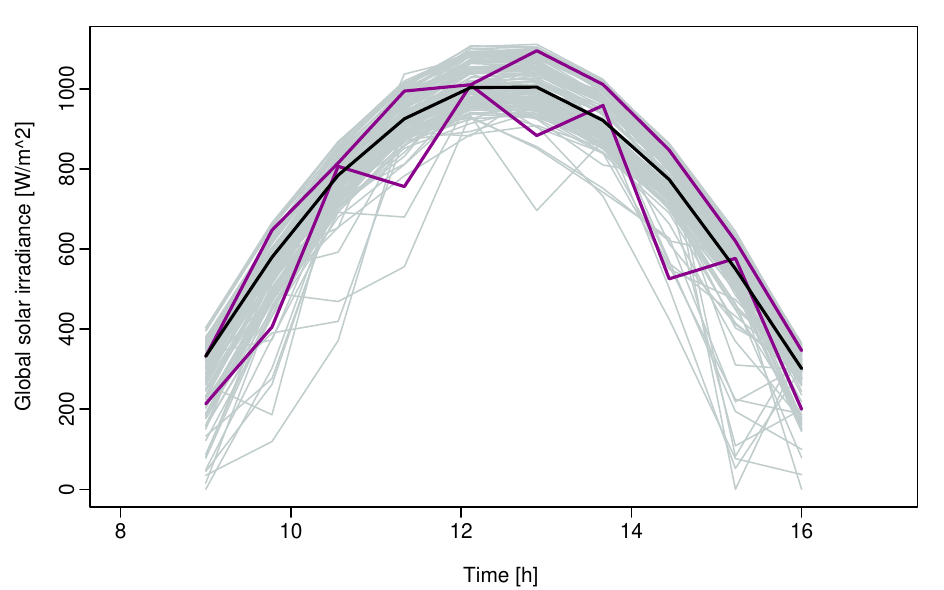}  
    \caption{Daily curves from 15 January 2019 to 15 June 2019. The top left panel represents the PV power with magnitude outliers flagged in blue. The top right panel represents the PV power with shape outliers flagged in magenta. The bottom left panel represents the global solar irradiance with magnitude outliers flagged in blue. The bottom right panel represents the global solar irradiance with shape outliers flagged in magenta.}
    \label{app:winter2019}
\end{figure}

\begin{figure}[!ht]
\centering
\includegraphics[width=75mm]{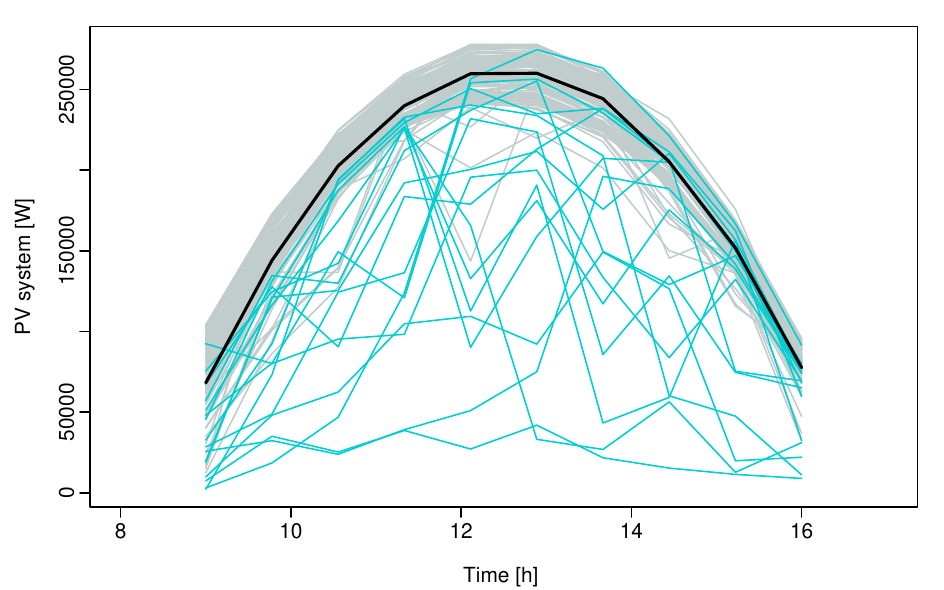}
\quad
\includegraphics[width=75mm]{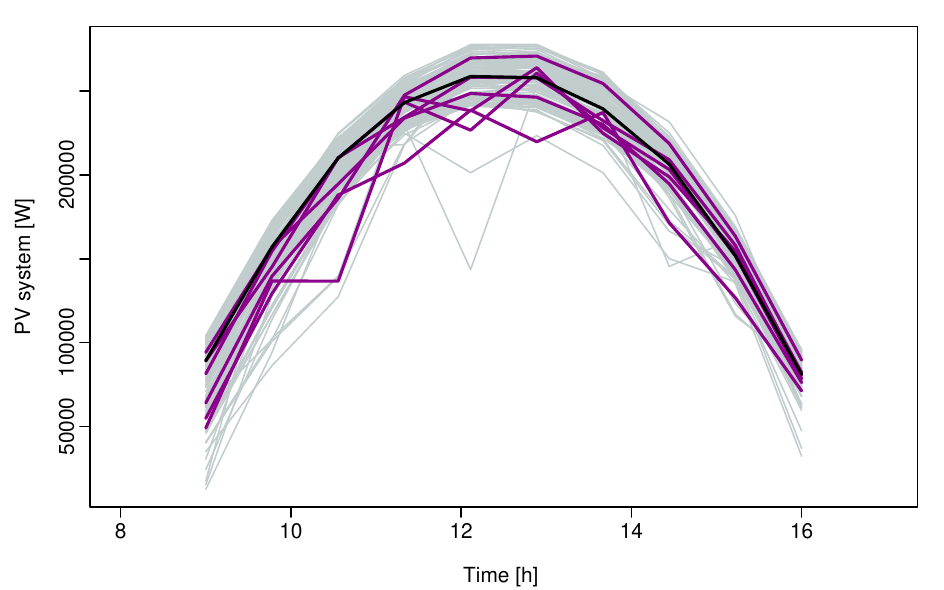}
\hspace{0mm}
\includegraphics[width=75mm]{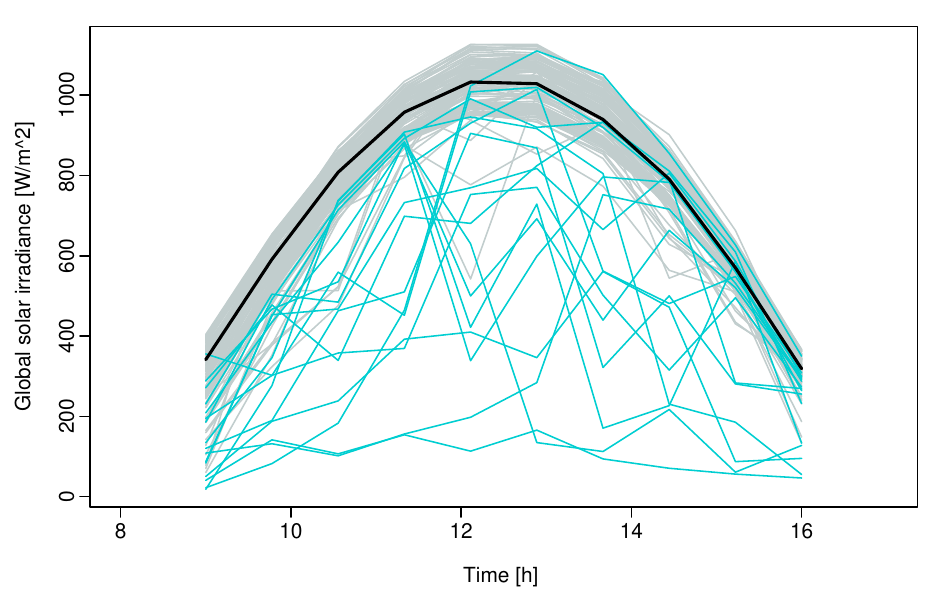}
\quad
\includegraphics[width=75mm]{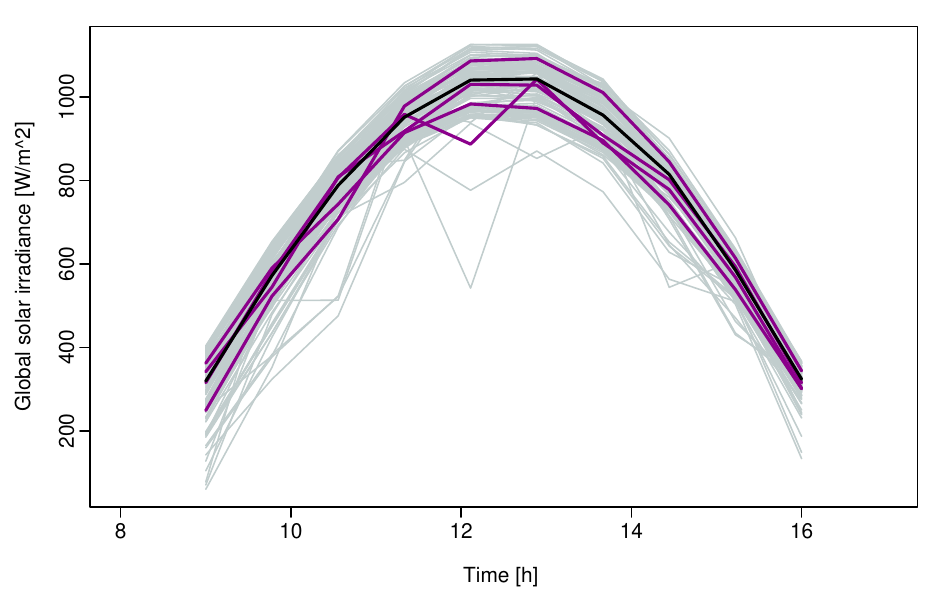}

    \caption{Daily curves from 15 January 2020 to 15 June 2020. The top left panel represents the PV power with magnitude outliers flagged in blue. The top right panel represents the PV power with shape outliers flagged in magenta. The bottom left panel represents the global solar irradiance with magnitude outliers flagged in blue. The bottom right panel represents the global solar irradiance with shape outliers flagged in magenta.}
    \label{app:winter2020}
\end{figure}

In this experiment, we apply the proposed univariate approach to daily curves for the two considered variables separately. Figures \ref{app:winter2019} and \ref{app:winter2020} consider daily curves from 15 January to 15 June in 2019 and 2020, respectively. In all cases, the median curve is highlighted in black.  The top left panel represents the daily curves of PV power; those curves flagged in blue are magnitude outliers detected by the functional boxplot. The top right panel shows the daily PV power curves after removing magnitude outliers. The curves in magenta are shape outliers detected by the proposed method. The bottom panels are the counterparts for global solar irradiance.  Note that only observed data were used in this study, with no introduction of artificial outliers.
 Figures  \ref{app:winter2019} and \ref{app:winter2020}  show the flagged outliers by the proposed method in magenta; the majority of these outliers are exclusively recognized by the proposed method. The reason for this is that such outliers are mostly fluctuations around the central region; these changes in shape are hard to capture by other methods. Those highlighted shape outliers present variability in the magnitude of the PV power and global solar irradiance, generating changes in the measurement around the median. Those changes around the median level indicate technical failures in the measurement sensors, making the measurements unreliable to interpret. Since both the  PV power values (order of $10^5$) and global solar irradiance values (order of $10^3$) are large in magnitude, detecting changes around the median becomes relevant. In practice, such anomalies are known as sensor faults, which are represented by small and quick changes within the specification band of the respective variables \citep{mellit2018fault}. Such anomalies are characterized by having small deviations from the general pattern in the PV system as well as small drifts at random time points.

 \medskip
  Note that the grey profiles in the right panels of Figures~\ref{app:winter2019} and \ref{app:winter2020} exhibit significant deviations from the median but do not fall within our outlier classification. These grey profiles are characterized by isolated data points that deviate from the typical profile. However, they do not align with the stringent criteria applied in our outlier identification method, especially when dealing with hourly data. Higher-resolution data is imperative to achieve enhanced sensitivity in outlier detection as it provides a more intricate perspective of system behavior, facilitating the identification of subtle deviations. It is essential to emphasize that our present analysis is based on hourly data, and the level of detail inherent in this data granularity may impact the sensitivity of our outlier detection methodology. 

 In practice, several types of anomalies can degrade the productivity of a PV plant, including partial shading, short circuits, and dust accumulation on the PV modules~\citep{mellit2018fault}. For instance, if many PV modules are short-circuited, then the PV power will significantly decrease. Hence, a potential application of this approach is to detect anomalies in PV systems.

\begin{figure}[!ht]
\centering
\includegraphics[width=75mm]{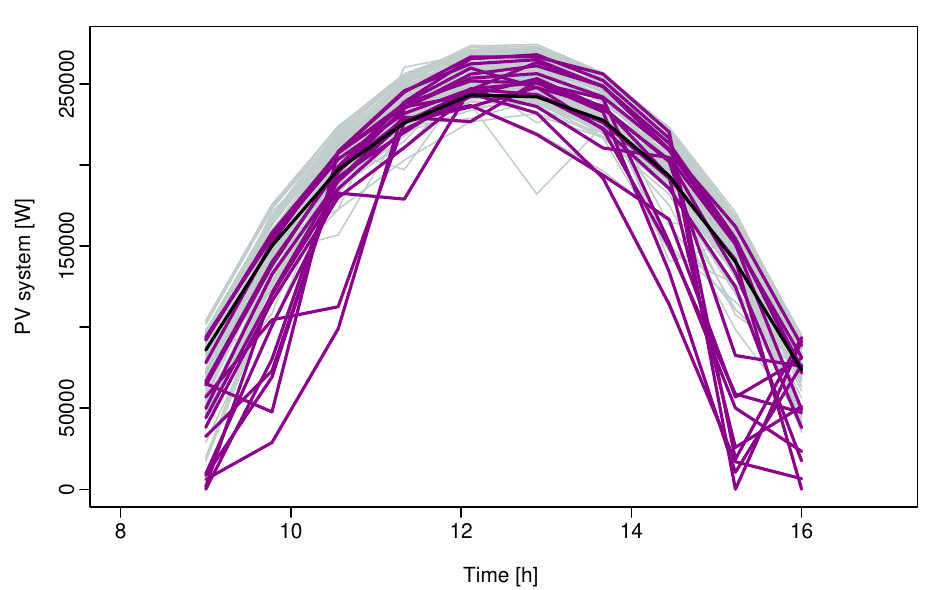}
\quad
\includegraphics[width=75mm]{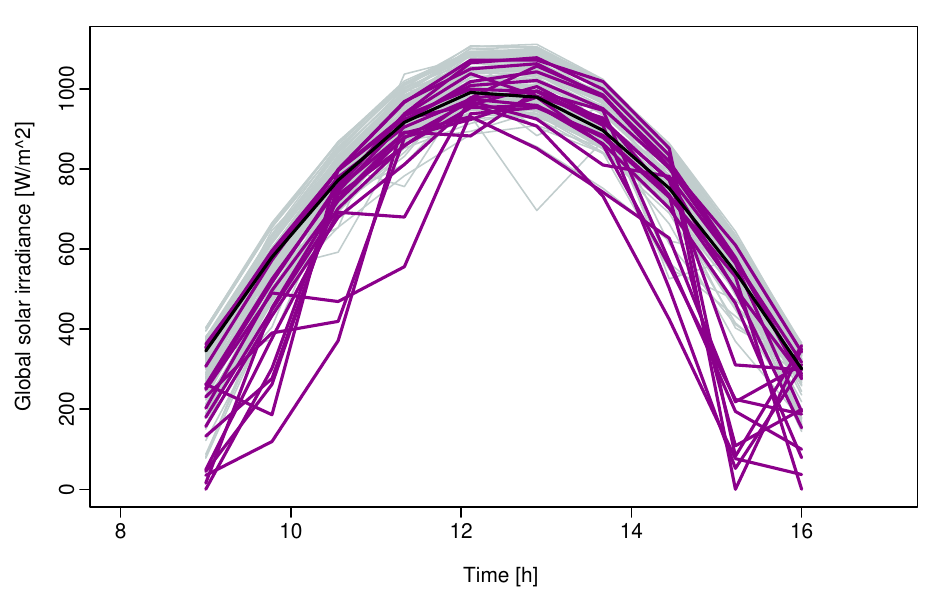}
\hspace{0mm}
\includegraphics[width=75mm]{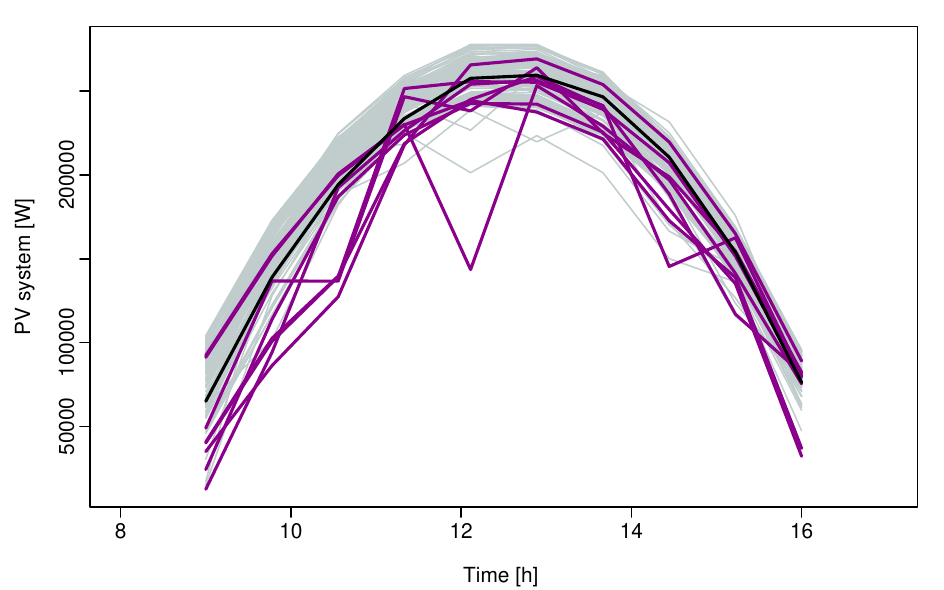}
\quad
\includegraphics[width=75mm]{msplot_SI1.pdf}
\caption{Shape outliers detected by MSplot after removing magnitude outliers using the functional boxplot. The left column displays PV power for 2019 (top) and 2020 (bottom), while the corresponding global solar irradiance is shown in the right column, with flagged shape outliers highlighted in magenta.}
    \label{app:msplot}
\end{figure}

As previously indicated, the emphasis is on recognizing shape outliers, particularly those with frequent fluctuations around the functional median level. In Figure \ref{app:msplot}, a comparison is presented regarding the results of identifying shape outliers using the most recent functional outlier detection method, the MS plot by \cite{Dai2018}, as discussed in the simulation section. The left column represents the PV power, while the right column represents the solar irradiance. It can be observed that the MS plot identifies some additional curves as outliers, most of which exhibit deviations from the central but are unable to detect those with frequent variations around the median.

In conclusion, the MS plot and our approach exhibit complementary capabilities in outlier detection, as each method identifies outliers that the other may miss. Leveraging both approaches can provide a more comprehensive and robust strategy for recognizing a broader range of outlier patterns in functional data.

\subsection{Multivariate}

\begin{figure}[!ht]
\centering
  \includegraphics[width=70mm]{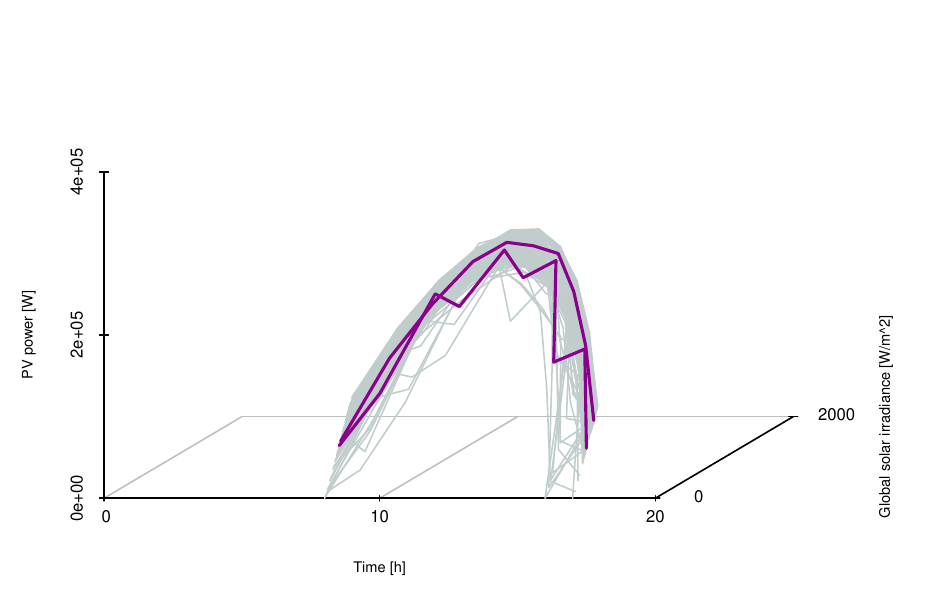}
  \quad
  \includegraphics[width=70mm]{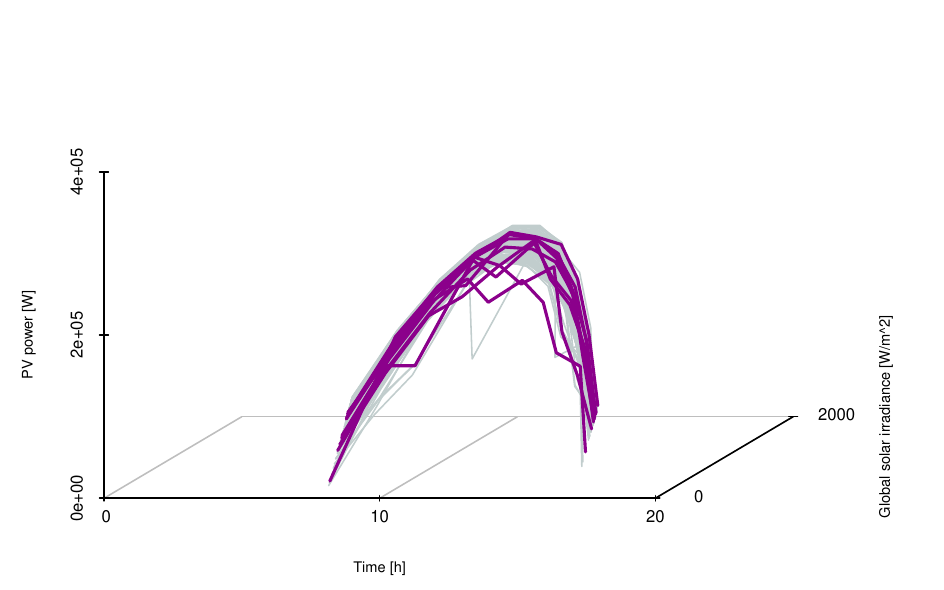}
   \caption{The left panel represents the bivariate functional data with the two variables PV power and global solar irradiance from 15 January 2019 to 15 June 2019, and the right panel represents the counterpart in 2020. Those curves flagged in magenta are shape outliers by the PD of the pointwise simplicial depth.}
    \label{app:bivariate}
\end{figure}

In Section~\ref{app:univ}, outliers are detected independently without considering the correlations between PV and global solar irradiance. In this section, we treat PV power and global solar irradiance as bivariate functional data and detect bivariate functional outliers. This expanded perspective allows us to detect outliers that may have been overlooked in the univariate analysis. Multivariate functional outlier detection considers the joint behavior of multiple variables simultaneously. In contrast, univariate methods treat each variable independently, potentially missing out on important interdependencies between variables.  This is particularly relevant because, in real-world scenarios, the performance of a PV system is often influenced by both PV power and solar irradiance in a correlated manner.  For example, a sudden drop in solar irradiance may lead to an unexpected decrease in PV power, and this relationship is crucial to understanding system anomalies. Furthermore, multivariate functional outlier detection provides richer information about the nature of outliers. It not only identifies that an outlier exists but can also describe how the joint behavior of variables differs from the norm. This detailed characterization is valuable for further analysis and decision-making.  In applications where system performance is influenced by multiple variables, such as in the case of PV systems where PV power and solar irradiance are interconnected, multivariate methods facilitate a more holistic approach to decision-making. They allow for a comprehensive assessment of how anomalies in one variable may impact others, aiding in better-informed responses.
Figure \ref{app:bivariate} represents the bivariate curves for the two time periods considered in the univariate case. The left panel shows the bivariate functional data with the two variables PV power and global solar irradiance from 15 January to 15 June 2019 after magnitude outliers are removed. Those curves flagged in magenta are shape outliers detected by the proposed method based on the sample correlation of the pairwise simplicial depth. The corresponding for 2020 is in the right panel.  \\

\begin{figure}[H]
\centering
  \includegraphics[width=70mm]{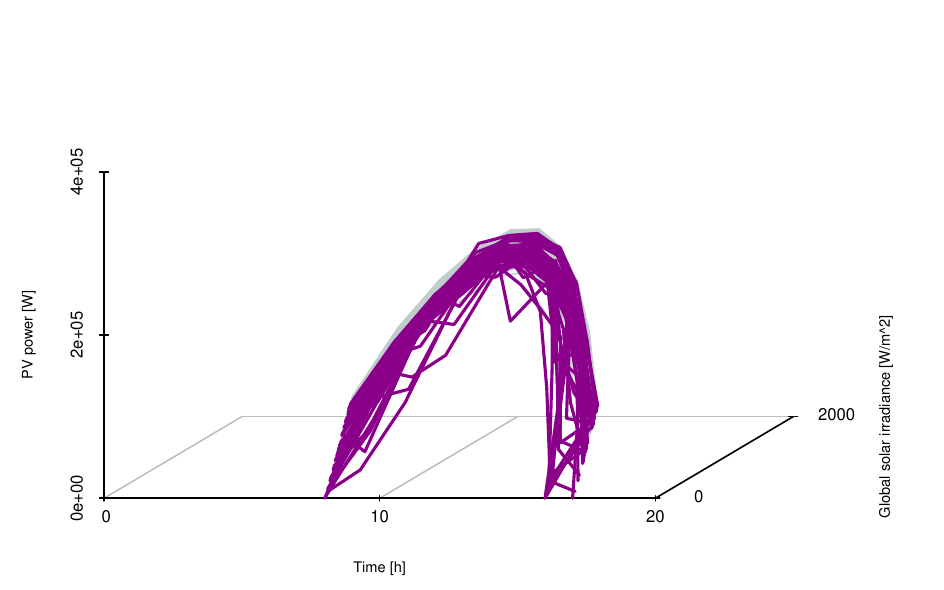}
  \quad
  \includegraphics[width=70mm]{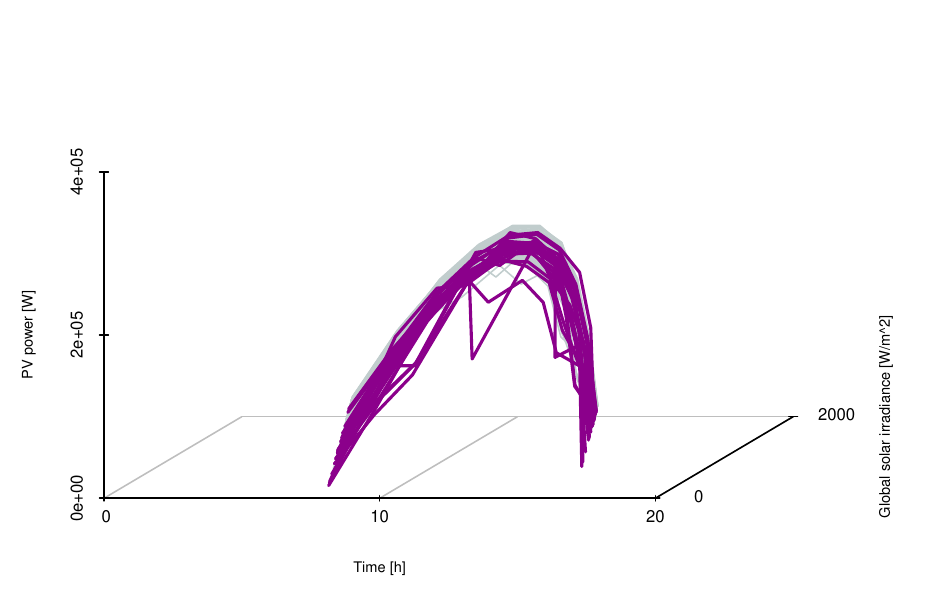}
\caption{ The left panel represents the bivariate functional data with two variables: PV power and global solar irradiance from 15 January 2019 to 15 June 2019, and the right panel represents the counterpart in 2020. The curves flagged in magenta are shape outliers detected by the MS plot.}   
    \label{app:msplot_bivariate}
\end{figure}

In the multivariate context, MS plot cannot detect some of the shape outliers depicted in Figure \ref{app:bivariate}. In practice, the two approaches could complement each other. Figure \ref{app:msplot_bivariate} displays the MS plot's detection of bivariate functional outliers. As discussed in the univariate case, the MS plot method often identifies more curves as outliers, leading to higher false positive rates. However, it still may not capture certain outliers characterized by frequent changes around the median level, where the pairwise depth method performs better.

\section{Discussion}\label{sec:Discussion}

In this paper, we proposed using pointwise depth to analyze anomalous observations in functional datasets. This proposal evidences different advantages in its application, and it is a promising alternative in comparison with the existing options. Along with the univariate distribution of the pointwise depth, the pairwise depth allows for better visualization of both magnitude and shape outliers.
Through intensive simulation studies, we show that in comparison to the most popular alternatives in the recent literature, our method has always detected outliers with frequent changes around the functional median. As part of the proposed outlier detection algorithms, we also proposed very useful visualization tools by observing both the pointwise depth and pairwise depth distributions for detecting shape outliers. The sample correlation of the pairwise distribution of the pointwise depth is a simple but very powerful alternative to identify shape outliers. Finally, using the pointwise data in the univariate case, it allows an easy and computationally achievable extension to multivariate functional observations. 

\medskip
In this study, we applied the proposed outlier detection method to monitor data from PV systems. In future work, it will be interesting to assess this approach by monitoring data from other renewable energy systems, such as wind turbines.

\bibliographystyle{chicago}
\bibliography{Paper1}

\begin{thebibliography}{}

\bibitem[\protect\citeauthoryear{Alam, Khan, Johnson, and Flicker}{Alam
  et~al.}{2015}]{alam2015comprehensive}
Alam, M.~K., F.~Khan, J.~Johnson, and J.~Flicker (2015).
\newblock A comprehensive review of catastrophic faults in {PV} arrays: types,
  detection, and mitigation techniques.
\newblock {\em IEEE Journal of Photovoltaics\/}~{\em 5\/}(3), 982--997.

\bibitem[\protect\citeauthoryear{Araveti, Quintana, Kairisa, Mutule, Adriazola,
  Sweeney, and Carroll}{Araveti et~al.}{2022}]{araveti2022wind}
Araveti, S., C.~A. Quintana, E.~Kairisa, A.~Mutule, J.~P.~S. Adriazola,
  C.~Sweeney, and P.~Carroll (2022).
\newblock {Wind Energy Assessment for Renewable Energy Communities}.
\newblock {\em Wind\/}~{\em 2\/}(2), 325--347.

\bibitem[\protect\citeauthoryear{Arribas-Gil and Romo}{Arribas-Gil and
  Romo}{2014}]{outliergram}
Arribas-Gil, A. and J.~Romo (2014).
\newblock Shape outlier detection and visualization for functional data: {T}he
  outliergram.
\newblock {\em Biostatistics\/}~{\em 15}, 603--619.

\bibitem[\protect\citeauthoryear{Barnett}{Barnett}{1976}]{Barnett1976}
Barnett, V. (1976).
\newblock {The Ordering of Multivariate Data}.
\newblock {\em Journal of the Royal Statistical Society. Series A
  (General)\/}~{\em 139\/}(3), 318--355.

\bibitem[\protect\citeauthoryear{Chakraborty and Chaudhuri}{Chakraborty and
  Chaudhuri}{2014}]{Chakraborty}
Chakraborty, A. and P.~Chaudhuri (2014).
\newblock On data depth in infinite dimensional spaces.
\newblock {\em Annals of the Institute of Statistical Mathematics\/}~{\em 66},
  303--324.

\bibitem[\protect\citeauthoryear{Cuevas, Febrero, and Fraiman}{Cuevas
  et~al.}{2007}]{Cuevas}
Cuevas, A., M.~Febrero, and R.~Fraiman (2007).
\newblock {Robust estimation and classification for functional data via
  projection-based depth notions}.
\newblock {\em Computational Statistics\/}~{\em 22\/}(3), 481--496.

\bibitem[\protect\citeauthoryear{Dai and Genton}{Dai and
  Genton}{2018}]{Dai2018}
Dai, W. and M.~G. Genton (2018).
\newblock Multivariate {F}unctional {D}ata {V}isualization and {O}utlier
  {D}etection.
\newblock {\em Journal of Computational and Graphical Statistics\/}~{\em 27},
  923--934.

\bibitem[\protect\citeauthoryear{Dai and Genton}{Dai and Genton}{2019}]{msplot}
Dai, W. and M.~G. Genton (2019).
\newblock Directional outlyingness for multivariate functional data.
\newblock {\em Computational Statistics and Data Analysis\/}~{\em 131}, 50--65.

\bibitem[\protect\citeauthoryear{Dai, Mrkvička, Sun, and Genton}{Dai
  et~al.}{2020}]{Daitransf}
Dai, W., T.~Mrkvička, Y.~Sun, and M.~G. Genton (2020).
\newblock Functional outlier detection and taxonomy by sequential
  transformations.
\newblock {\em Computational Statistics \& Data Analysis\/}~{\em 149}.

\bibitem[\protect\citeauthoryear{Dutta, Ghosh, and Chaudhuri}{Dutta
  et~al.}{2011}]{Dutta}
Dutta, S., A.~K. Ghosh, and P.~Chaudhuri (2011).
\newblock {Some intriguing properties of Tukey’s half-space depth}.
\newblock {\em Bernoulli\/}~{\em 17\/}(4), 1420 -- 1434.

\bibitem[\protect\citeauthoryear{Euán, Sun, and Reich}{Euán
  et~al.}{2022}]{Euan2022}
Euán, C., Y.~Sun, and B.~J. Reich (2022).
\newblock Statistical analysis of multi-day solar irradiance using a threshold
  time series model.
\newblock {\em Environmetrics\/}~{\em 33\/}(3), e2716.

\bibitem[\protect\citeauthoryear{Febrero, Galeano, and
  Gonzalez-Manteiga}{Febrero et~al.}{2008}]{Febrero2008}
Febrero, M., P.~Galeano, and W.~Gonzalez-Manteiga (2008).
\newblock Outlier detection in functional data by depth measures, with
  application to indentify abnormal {NO}x levels.
\newblock {\em Environmetrics\/}, 331345.

\bibitem[\protect\citeauthoryear{Ferraty and Vieu}{Ferraty and
  Vieu}{2006}]{Ferraty}
Ferraty, F. and P.~Vieu (2006).
\newblock {\em Nonparametric {F}unctional {D}ata {A}nalysis: {T}heory and
  {P}ractice (Springer Series in Statistics)}.
\newblock Berlin, Heidelberg: Springer-Verlag.

\bibitem[\protect\citeauthoryear{Fraiman and Muniz}{Fraiman and
  Muniz}{2001}]{Fraiman}
Fraiman, R. and G.~Muniz (2001).
\newblock Trimmed means for functional data.
\newblock {\em Test\/}~{\em 10}, 419--440.

\bibitem[\protect\citeauthoryear{Garoudja, Harrou, Sun, Kara, Chouder, and
  Silvestre}{Garoudja et~al.}{2017}]{Garoud2017}
Garoudja, E., F.~Harrou, Y.~Sun, K.~Kara, A.~Chouder, and S.~Silvestre (2017).
\newblock Statistical fault detection in photovoltaic systems.
\newblock {\em Solar Energy\/}~{\em 150}, 485--499.

\bibitem[\protect\citeauthoryear{Genton and Ma}{Genton and
  Ma}{1999}]{Genton1999}
Genton, M.~G. and Y.~Ma (1999).
\newblock Robustness properties of dispersion estimators.
\newblock {\em Statistics \& Probability Letters\/}~{\em 44}, 343--350.

\bibitem[\protect\citeauthoryear{Gneiting, Kleiber, and Schlather}{Gneiting
  et~al.}{2010}]{matern}
Gneiting, T., W.~Kleiber, and M.~Schlather (2010).
\newblock Matérn {C}ross-{C}ovariance {F}unctions for {M}ultivariate {R}andom
  {F}ields.
\newblock {\em Journal of the American Statistical Association\/}~{\em
  105\/}(491), 1167--1177.

\bibitem[\protect\citeauthoryear{Huang and Sun}{Huang and Sun}{2019}]{TVD}
Huang, H. and Y.~Sun (2019).
\newblock A {D}ecomposition of {T}otal {V}ariation {D}epth for {U}nderstanding
  {F}unctional {O}utliers.
\newblock {\em Technometrics\/}~{\em 61}, 445--458.

\bibitem[\protect\citeauthoryear{Hubert, Rousseeuw, and Segaert}{Hubert
  et~al.}{2015}]{Hubert2015}
Hubert, M., P.~J. Rousseeuw, and P.~Segaert (2015).
\newblock Multivariate functional outlier detection.
\newblock {\em Statistical Methods and Applications\/}~{\em 24}, 177--202.

\bibitem[\protect\citeauthoryear{Kleissl}{Kleissl}{2013}]{Kleissl2013}
Kleissl, J. (2013).
\newblock {\em {Solar Energy Forecasting and Resource Assessment}}.
\newblock Elsevier Science.

\bibitem[\protect\citeauthoryear{Liu}{Liu}{1990}]{simpl}
Liu, R.~Y. (1990).
\newblock {On a {N}otion of {D}ata {D}epth {B}ased on {R}andom {S}implices}.
\newblock {\em The Annals of Statistics\/}~{\em 18\/}(1), 405 -- 414.

\bibitem[\protect\citeauthoryear{Liu, Parelius, and Singh}{Liu
  et~al.}{1999}]{Liu1999}
Liu, R.~Y., J.~M. Parelius, and K.~Singh (1999).
\newblock Multivariate analysis by data depth: descriptive statistics, graphics
  and inference.
\newblock {\em The Annals of Statistics\/}~{\em 27}, 858.

\bibitem[\protect\citeauthoryear{Lopez-Pintado and Romo}{Lopez-Pintado and
  Romo}{2009}]{MBD}
Lopez-Pintado, S. and J.~Romo (2009).
\newblock On the concept of depth for functional data.
\newblock {\em Journal of the American Statistical Association\/}~{\em 104},
  718--734.

\bibitem[\protect\citeauthoryear{López-Pintado and Romo}{López-Pintado and
  Romo}{2011}]{halfregion}
López-Pintado, S. and J.~Romo (2011).
\newblock {A half-region depth for functional data}.
\newblock {\em Computational Statistics \& Data Analysis\/}~{\em 55\/}(4),
  1679--1695.

\bibitem[\protect\citeauthoryear{López-Pintado, Sun, Lin, and
  Genton}{López-Pintado et~al.}{2014}]{multi}
López-Pintado, S., Y.~Sun, J.~K. Lin, and M.~G. Genton (2014).
\newblock Simplicial band depth for multivariate functional data.
\newblock {\em Advances in Data Analysis and Classification\/}~{\em 8},
  321--338.

\bibitem[\protect\citeauthoryear{Ma and Genton}{Ma and
  Genton}{2000}]{Genton2000}
Ma, Y. and M.~G. Genton (2000).
\newblock Highly robust estimation of the autocovariance function.
\newblock {\em Journal of Time Series Analysis\/}~{\em 21\/}(6), 663--684.

\bibitem[\protect\citeauthoryear{Ma and Genton}{Ma and
  Genton}{2001}]{Genton2001}
Ma, Y. and M.~G. Genton (2001).
\newblock Highly {R}obust {E}stimation of {D}ispersion matrices.
\newblock {\em Journal of Multivariate Analysis\/}~{\em 78\/}(1), 11–36.

\bibitem[\protect\citeauthoryear{Mahalanobis}{Mahalanobis}{1936}]{mahalanobis}
Mahalanobis, P.~C. (1936).
\newblock On the generalized distance in statistics.
\newblock {\em Proceedings of the National Institute of Sciences
  (Calcutta)\/}~{\em 2}, 49--55.

\bibitem[\protect\citeauthoryear{Mallor, León, {De Boeck}, {Van Gulck},
  Meulders, and {Van der Meerssche}}{Mallor et~al.}{2017}]{Mallor2017}
Mallor, F., T.~León, L.~{De Boeck}, S.~{Van Gulck}, M.~Meulders, and B.~{Van
  der Meerssche} (2017).
\newblock A method for detecting malfunctions in {PV} solar panels based on
  electricity production monitoring.
\newblock {\em Solar Energy\/}~{\em 153}, 51--63.

\bibitem[\protect\citeauthoryear{Marcos, Marroyo, Lorenzo, Alvira, and
  Izco}{Marcos et~al.}{2011}]{Marcos2011}
Marcos, J., L.~Marroyo, E.~Lorenzo, D.~Alvira, and E.~Izco (2011).
\newblock {Power output fluctuations in large scale {PV} plants: One year
  observations with one second resolution and a derived analytic model}.
\newblock {\em Progress in Photovoltaics: Research and Applications\/}~{\em
  19\/}(2), 218--227.

\bibitem[\protect\citeauthoryear{Mat{\'e}rn}{Mat{\'e}rn}{1960}]{matern2013spatial}
Mat{\'e}rn, B. (1960).
\newblock {\em Spatial {V}ariation}, Volume~36.
\newblock Springer Science \& Business Media.

\bibitem[\protect\citeauthoryear{Mellit, Tina, and Kalogirou}{Mellit
  et~al.}{2018}]{mellit2018fault}
Mellit, A., G.~M. Tina, and S.~A. Kalogirou (2018).
\newblock Fault detection and diagnosis methods for photovoltaic systems: A
  review.
\newblock {\em Renewable and Sustainable Energy Reviews\/}~{\em 91}, 1--17.

\bibitem[\protect\citeauthoryear{Narisetty and Nair}{Narisetty and
  Nair}{2016}]{Narisetty2016}
Narisetty, N.~N. and V.~N. Nair (2016).
\newblock Extremal {D}epth for {F}unctional {D}ata and {A}pplications.
\newblock {\em Journal of the American Statistical Association\/}~{\em 111},
  1705--1714.

\bibitem[\protect\citeauthoryear{Psomopoulos, Ioannidis, Kaminaris, Mardikis,
  and Katsikas}{Psomopoulos et~al.}{2015}]{psomopoulos2015comparative}
Psomopoulos, C.~S., G.~C. Ioannidis, S.~D. Kaminaris, K.~D. Mardikis, and N.~G.
  Katsikas (2015).
\newblock A comparative evaluation of photovoltaic electricity production
  assessment software ({PVGIS}, {PVW}atts and {RETS}creen).
\newblock {\em Environmental Processes\/}~{\em 2\/}(1), 175--189.

\bibitem[\protect\citeauthoryear{Ramsay and Silverman}{Ramsay and
  Silverman}{2006}]{Ramsay}
Ramsay, J. and B.~Silverman (2006).
\newblock {\em Functional {D}ata {A}nalysis}.
\newblock Springer Series in Statistics. New York: Springer.

\bibitem[\protect\citeauthoryear{REN22.}{REN22.}{2022}]{REN21}
REN22. (2022).
\newblock Renewables 2022 global status report.
\newblock Technical report.

\bibitem[\protect\citeauthoryear{Serfling}{Serfling}{2002}]{Serfling2002}
Serfling, R. (2002).
\newblock {A} {D}epth {F}unction and a {S}cale {C}urve {B}ased on {S}patial
  {Q}uantiles.
\newblock In Y.~Dodge (Ed.), {\em Statistical Data Analysis Based on the
  L1-Norm and Related Methods}, Basel, pp.\  25--38. Birkh{\"a}user Basel.

\bibitem[\protect\citeauthoryear{Solórzano and Egido}{Solórzano and
  Egido}{2013}]{Solorzano2013}
Solórzano, J. and M.~Egido (2013).
\newblock {A}utomatic {f}ault diagnosis in {PV} systems with distributed
  {MPPT}.
\newblock {\em Energy Conversion and Management\/}~{\em 76}, 925--934.

\bibitem[\protect\citeauthoryear{Sun and Genton}{Sun and
  Genton}{2011}]{Sun2011}
Sun, Y. and M.~G. Genton (2011).
\newblock Functional {B}oxplots.
\newblock {\em Journal of Computational and Graphical Statistics\/}~{\em 20},
  316--334.

\bibitem[\protect\citeauthoryear{Sun and Genton}{Sun and
  Genton}{2012}]{AdjSun2012}
Sun, Y. and M.~G. Genton (2012).
\newblock Adjusted functional boxplots for spatio-temporal data visualization
  and outlier detection.
\newblock {\em Environmetrics\/}~{\em 23}, 54--64.

\bibitem[\protect\citeauthoryear{Sun, Genton, and Nychka}{Sun
  et~al.}{2012}]{comp}
Sun, Y., M.~G. Genton, and D.~W. Nychka (2012).
\newblock Exact fast computation of band depth for large functional datasets:
  How quickly can one million curves be ranked?
\newblock {\em Stat\/}~{\em 1\/}(1), 68--74.

\bibitem[\protect\citeauthoryear{{\v{S}}{\'u}ri, Huld, and
  Dunlop}{{\v{S}}{\'u}ri et~al.}{2005}]{vsuri2005pv}
{\v{S}}{\'u}ri, M., T.~A. Huld, and E.~D. Dunlop (2005).
\newblock {PV-GIS: a web-based solar radiation database for the calculation of
  PV potential in Europe}.
\newblock {\em International Journal of Sustainable Energy\/}~{\em 24\/}(2),
  55--67.

\bibitem[\protect\citeauthoryear{Tukey}{Tukey}{1975}]{tukey}
Tukey, J.~W. (1975).
\newblock Mathematics and the {P}icturing of {D}ata.
\newblock {\em Proceedings of the International Congress of Mathematicians,
  Vancouver, 1975\/}~{\em 2}, 523--531.

\bibitem[\protect\citeauthoryear{Ullah and Finch}{Ullah and
  Finch}{2013}]{ullah2013}
Ullah, S. and C.~F. Finch (2013).
\newblock Applications of functional data analysis: A systematic review.
\newblock {\em BMC medical research methodology\/}~{\em 13\/}(1), 1--12.

\bibitem[\protect\citeauthoryear{Zuo}{Zuo}{2003}]{Zuo2003}
Zuo, Y. (2003).
\newblock Projection-based depth functions and associated medians.
\newblock {\em The Annals of Statistics\/}~{\em 31}, 1460--1490.

\bibitem[\protect\citeauthoryear{Zuo and Serfling}{Zuo and
  Serfling}{2000}]{Zuo2000}
Zuo, Y. and R.~Serfling (2000).
\newblock General notions of statistical depth function.
\newblock {\em The Annals of Statistics\/}~{\em 28}, 461--482.

\end{thebibliography}

\clearpage
\section{Appendix}\label{Appendix}
\subsection{Simulation study for dependence shape outliers with different parameters in the covariance function}\label{Appendix_1}

Regarding the sensibility of our method regarding the characterizations of the covariance function of the outlying curves, we designed some additional simulations for the dependence shape outlier (Model 1) in which the general exponential covariance function of the form $\gamma(s,t)= k*\exp\{-(1/c)*|s-t|^{\mu}\}$ is analyzed by varying the two main parameters $k=(2,4,6)$ and $\mu=(0.1,0.5,0.7)$.

\begin{table}[ht]
\centering
\begin{tabular}{rrrrrrrrrr}
  \hline
 $(k,\mu):=$& $(2,0.1)$ & $(4,0.1)$ & $(6,0.1)$ & $(2,0.5)$ & $(4,0.5)$ & $(6,0.5)$ & $(2,0.7)$ & $(4,0.7)$ & $(6,0.7)$ \\ 
  \hline
TPR & 100.00 & 100.00 & 100.00 & 68.64 & 79.43 & 85.40 & 27.79 & 37.77 & 40.89 \\ 
  sd & 0.00 & 0.00 & 0.00 & 19.05 & 17.85 & 15.47 & 16.10 & 16.90 & 19.99 \\ 
  FPR & 2.67 & 2.72 & 2.78 & 2.17 & 2.20 & 2.17 & 3.07 & 2.64 & 2.42 \\ 
  sd & 1.45 & 1.44 & 1.43 & 1.63 & 1.64 & 1.68 & 1.79 & 1.68 & 1.61 \\ 
   \hline
\end{tabular}
\caption{Sensibility analysis of the parameters of the covariance function in Model 1 described in Section~\ref{sec:Simulation}. TPR refers to the average across 500 experiments, sd denotes the corresponding standard deviation.}
\label{SA_mu_k}
\end{table}

Table \ref{SA_mu_k} shows that as $\mu$ increases from $0.1$ to $0.7$, outlier models become increasingly similar to non-outlier models. When $k=2$, the TPR drops from $100\%$ to $27.79\%$. For $k=6$, despite a large variance, the TPR only decreases to $40.89\%$.

\subsection{Sensitivity analysis for the shape outlier models with different contamination rates $\theta$}\label{Appendix_2}

We implemented some simulations by increasing the contamination rate $\theta$ by taking different values $\theta=(0.1,0.15,0.2,0.24,0.30)$. From the point of view of the TPR in Table \ref{TPR}, on average as the contamination rates increase the mean TPR decreases, but its variability increases. From the FPR perspective in Table \ref{FPR}, on average as the contamination rate increases, the FPR decreases as well as its variability.

    \begin{table}[!ht]
\centering
\begin{tabular}{rrrrrrrrrrr}
  \hline
  & \multicolumn{2}{c}{Model 1} &\multicolumn{2}{c}{Model 2}&
  \multicolumn{2}{c}{Model 3} & \multicolumn{2}{c}{Model 4} &\multicolumn{2}{c}{Model 5}\\
  \hline
   & TPR & sd & TPR & sd & TPR & sd & TPR & sd & TPR & sd \\
     \hline
$\theta=0.1$  & 100.00 & 0.00 & 99.28 & 2.98 & 99.54 & 2.34 & 99.59 & 2.07 & 99.89 & 1.08 \\ 
  $\theta=0.15$  & 98.09 & 11.58 & 92.21 & 14.71 & 97.26 & 9.54 & 97.11 & 9.27 & 97.74 & 9.30 \\ 
  $\theta=0.2$  & 82.00 & 35.20 & 71.63 & 27.40 & 82.98 & 30.67 & 82.79 & 30.90 & 83.21 & 32.16 \\ 
  $\theta=0.25$ & 42.04 & 44.76 & 41.70 & 25.78 & 45.91 & 42.50 & 46.05 & 42.60 & 45.76 & 43.99 \\ 
  $\theta=0.3$ & 10.29 & 23.54 & 23.32 & 14.30 & 12.24 & 24.79 & 12.53 & 25.08 & 11.95 & 25.26 \\ 
   \hline
\end{tabular}
\caption{TPR variation with different contamination rates. TPR refers to the average across 500 experiments, sd denotes the corresponding standard deviation }
\label{TPR}
\end{table}

\begin{table}[!ht]
\centering
\begin{tabular}{rrrrrrrrrrr}
  \hline
   & \multicolumn{2}{c}{Model 1} &\multicolumn{2}{c}{Model 2}&
  \multicolumn{2}{c}{Model 3} & \multicolumn{2}{c}{Model 4} &\multicolumn{2}{c}{Model 5}\\
  \hline
  & FPR & sd & FPR & sd & FPR & sd & FPR & sd & FPR & sd \\ 
  \hline
$\theta=0.1$ & 2.81 & 1.54 & 2.41 & 1.41 & 2.67 & 1.50 & 2.64 & 1.51 & 2.67 & 1.48 \\ 
  $\theta=0.15$ & 1.95 & 1.25 & 1.41 & 1.29 & 1.69 & 1.25 & 1.70 & 1.26 & 1.78 & 1.24 \\ 
  $\theta=0.2$ & 1.13 & 1.01 & 0.58 & 0.96 & 0.84 & 1.04 & 0.85 & 0.98 & 0.96 & 1.03 \\ 
  $\theta=0.25$ & 0.36 & 0.71 & 0.17 & 0.54 & 0.27 & 0.69 & 0.27 & 0.71 & 0.33 & 0.72 \\ 
  $\theta=0.3$ & 0.04 & 0.26 & 0.15 & 0.58 & 0.04 & 0.26 & 0.04 & 0.27 & 0.05 & 0.30 \\ 
   \hline
\end{tabular}
\caption{FPR variation with different contamination rates. FPR refers to the average across 500 experiments, sd denotes the corresponding standard deviation}
\label{FPR}
\end{table}

\end{document}